\documentclass{aa}  

\usepackage{graphicx}
\usepackage{txfonts}
\usepackage{hyperref}
\begin{document}

   \title{The sharpest view on the high-mass star-forming region S255IR}

   \subtitle{Near-InfraRed Adaptive Optics Imaging on the Outbursting Source NIRS3}

   \author{R. Fedriani
          \inst{1,2}
          \and A. Caratti o Garatti\inst{3}
          \and
          R. Cesaroni\inst{4}
          \and
          J. C. Tan\inst{2,5}
          \and
          B. Stecklum\inst{6}
          \and
          L. Moscadelli\inst{4}
          \and
          M. Koutoulaki\inst{7}
          \and
          G. Cosentino\inst{2}
          \and
          M. Whittle\inst{5}
          }

   \institute{Instituto de Astrof\'isica de Andaluc\'ia, CSIC, Glorieta de la Astronom\'ia s/n, E-18008 Granada, Spain\\
              \email{fedriani@iaa.es}
            \and
            Dept. of Space, Earth \& Environment, Chalmers University of Technology, 412 93 Gothenburg, Sweden    
            \and
            INAF - Osservatorio Astronomico di Capodimonte, salita Moiariello 16, 80131, Napoli, Italy
            % \and
            % Dublin Institute for Advanced Studies, Astronomy \& Astrophysics Section, 31 Fitzwilliam Place, Dublin 2, Ireland
            \and INAF, Osservatorio Astrofisico di Arcetri, Largo Fermi 5, I-50125 Firenze, Italy
            % \and Department of Space, Earth \& Environment, Chalmers University of Technology, 412 93 Gothenburg, Sweden
            \and
            Department of Astronomy, University of Virginia, Charlottesville, Virginia 22904, USA
            \and
            Th\"uringer Landessternwarte Tautenburg, Sternwarte 5, 07778 Tautenburg, Germany 
            \and
            School of Physics and Astronomy, University of Leeds, Leeds LS2 9JT, UK
            }

   \date{Received 27 April 2023 / Accepted 23 June 2023}

% \abstract{}{}{}{}{} 
% 5 {} token are mandatory
 
  \abstract
  % context heading (optional)
  % {} leave it empty if necessary  
   {Massive stars have an impact on their surroundings from early in their formation until the end of their lives. However, very little is known about their formation. Episodic accretion may play a crucial role, but observations of these events have only been reported towards a handful of massive protostars.}
  % aims heading (mandatory)
   {We aim to investigate the outburst event from the high-mass star-forming region S255IR where recently the protostar NIRS3 underwent an accretion outburst. We follow the evolution of this source both in photometry and morphology of its surroundings.}
  % methods heading (mandatory)
   {We perform near-infrared adaptive optics observations on the S255IR central region using the Large Binocular Telescope in the K$_{\rm s}$ broad-band and the H$_2$ and Br$\gamma$ narrow-band filters with an angular resolution of $\sim0\farcs06$, close to the diffraction limit.}
  % results heading (mandatory)
   {We discover a new near-infrared knot north-east from NIRS3 that we interpret as a jet knot that was ejected during the last accretion outburst and observed in the radio regime as part of a follow-up after the outburst. We measure a mean tangential velocity for this knot of $450\pm50\,\mathrm{km\,s^{-1}}$. We analyse the continuum-subtracted images from H$_2$ which traces jet shocked emission, and Br$\gamma$ which traces scattered light from a combination of accretion activity and UV radiation from the central massive protostar. We observe a significant decrease in flux at the location of NIRS3, with K=13.48\,mag being the absolute minimum in the historic series.}
  % conclusions heading (optional), leave it empty if necessary 
   {Our observations strongly suggest a scenario where the episodic accretion is followed by an episodic ejection response in the near-infrared, as it was seen in the earlier radio follow-up. The 30 years of $\sim2\,\mu{\rm m}$ photometry suggests that NIRS3 might have undergone another outburst in the late 1980s, being the first massive protostar with such evidence observed in the near-infrared.}


   \maketitle
%
%-------------------------------------------------------------------

\section{Introduction}\label{sect:introduction}

%Overview figure in Ks band and RGB
\begin{figure*}[!htb]
        \includegraphics[width=0.49\textwidth]{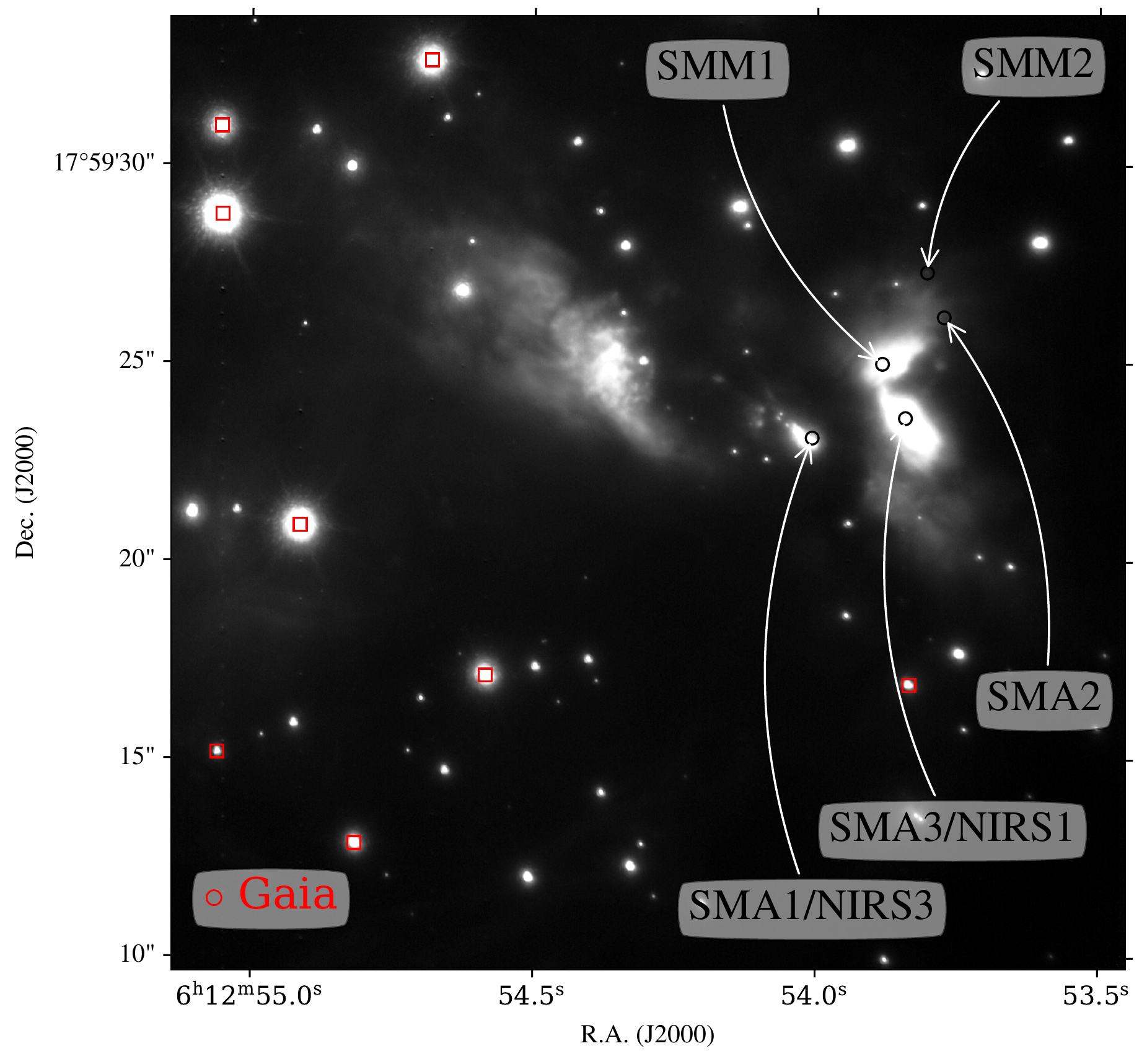}
        \includegraphics[width=0.49\textwidth]{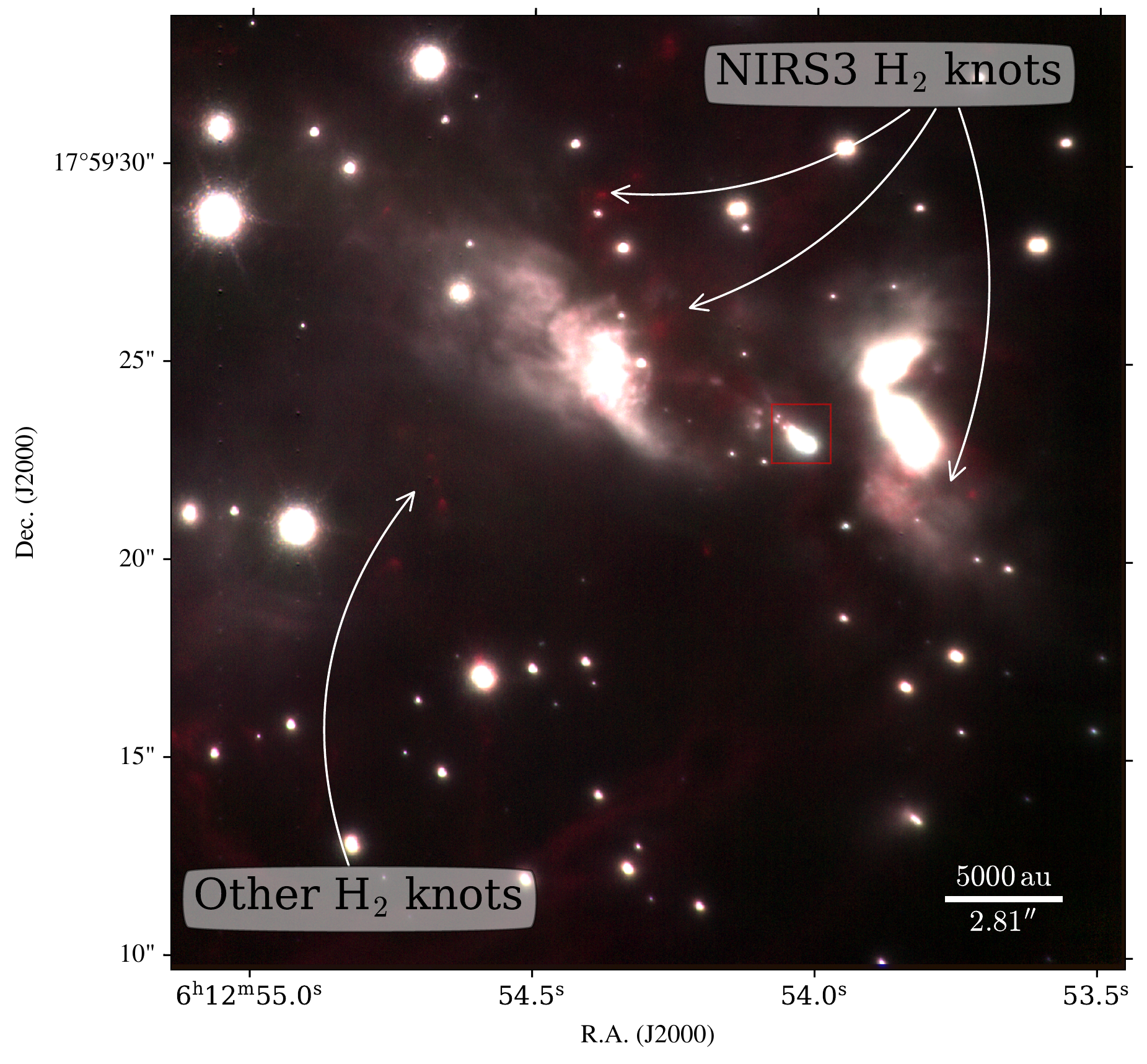}

  \caption{LBT/LUCI images for the S255IR star forming region. Left panel: AO-assisted K$_{\rm s}$ band image for the S255IR region displayed in linear scale. The FWHM at the position of the AO guide star is $\sim0\farcs06$ (see Appendix\,\ref{sect:appendix_soul} for more details). Black circles represent sources reported in Zichenko et al. (2020), whereas red squares represent Gaia sources used for astrometric solution. Right panel: RGB AO-assisted image with red H$_2$, green Br$\gamma$, and blue K$_{\rm s}$. The red box indicates the FoV shown in Figure\,\ref{fig:S255_multiband}.}
     \label{fig:continuum_image}
\end{figure*}

Massive stars strongly influence a vast range of astrophysical domains, from reionisation of the Universe, to formation of galaxies, to regulation of the star formation in molecular clouds \citep[see, e.g.,][]{tan2014}. Their formation, however, remains poorly understood. Nonetheless, the two key ingredients for forming stars, i.e. accretion disks and outflows, seem ubiquitous across a large mass range \citep{beltran2016,bally2016}. It has then been suggested that massive star formation may proceed as a scaled-up version of their lower mass counterparts. Furthermore, it has been observed in the low-mass regime that star formation may proceed episodically as opposed to a steadily \citep[see][for a recent review]{fischer2022}. Low-mass protostars may gather a significant amount of mass via episodic accretion. The latter manifests as accretion outbursts that can last from hours to decades and induce variability both photometrically and spectroscopically. A natural consequence of accretions disks is the ejection of outflows \citep{cabrit2007}. Their knotty structure also suggests episodic events in the formation of protostars \citep{zinnecker1998}. These episodic events, both in the form of accretion and ejection, seem to be relatively common in the low-mass regime \citep{fischer2022}. Recently, there have been a few high-mass young stellar object outbursts reported in the literature that may resemble the episodic events observed in their low-mass counterparts \citep{caratti2017nature,hunter2017,stecklum2021,chen2021}. This relatively new phenomenon is still unexplored and therefore further analysis is needed to identify the physical properties of episodic accretion/outflow events in high-mass systems.

The region S255IR represents a special case when referring to episodic events in high-mass young stellar objects (HMYSOs). The overall S255 region was first observed in the near-infrared (NIR) as part of the infrared survey by \citet{strom1976} using the 1.3 m telescope at Kitt Peak National Observatory. First reported by \citet{evans1977}, S255IR is located between the regions S255 and S257 in the cluster of optically visible HII regions S254-S258. S255IR is a high-mass star-forming region discovered by means of OH and H$_2$O maser observations \citep{turner1971,lo1975} and it is located at a distance of $1.78^{+0.12}_{-0.11}$\,kpc \citep{burns2016}. Observations of this region by \citet{howard1997} (who called it S255-2) and \citet{miralles1997} (who referred to it as S255IR) in the NIR at higher angular resolution, revealed a complex system with about 50-80 NIR sources. The main two young stellar objects (YSOs) in the nomenclature of \citet{miralles1997} (which we will follow in this paper) are NIRS1 \citep[denoted IRS 1a in][]{howard1997} and NIRS3 \citep[denoted IRS 1b in][]{howard1997}. \citet{tamura1991} performed NIR polarimetric observations on S255 and concluded that NIRS3 was the most likely illuminating source of the bipolar nebula. S255IR regained attention when \citet{fujisawa2015} reported a 6.7\,GHz class II methanol maser flare at the position of NIRS3 \citep[called SMA1 in][]{zinchenko2020}. Further NIR follow-up of this maser flare confirmed the outbursting nature of NIRS3 \citep{caratti2017nature}. The authors observed an increase in luminosity in both the H ($\sim1.6\,\mu{\rm m}$) and K ($\sim2.2\,\mu{\rm m}$) bands by $>2$\,mag deriving an enhanced accretion of $\sim3\times10^{-3}\,M_\odot\,\mathrm{yr^{-1}}$. A subsequent set of observations of NIRS3 in the radio regime with the Karl G. Jansky Very Large Telescope (VLA) revealed an exponential increase of the radio flux as a consequence of the accretion outburst \citep{cesaroni2018}.

Here, we report on new NIR adaptive optics assisted observations on the S255IR region with an angular resolution of $\sim0\farcs06$. The Letter is organised as follows: Section\,\ref{sect:observations} details the observations taken and the data reduction performed, Section\,\ref{sect:results} presents the results and discusses the main findings, and Section\,\ref{sect:conclusions} summarises the main conclusions.

%https://arxiv.org/pdf/2010.09199.pdf

% VVV YSO variability:
% https://arxiv.org/abs/2203.08681

%--------------------------------------------------------------------
\section{Observations and Data Reduction}\label{sect:observations}

Observations were taken on the 13th of February 2022 with the Large Binocular Telescope (LBT), in particular the SX telescope with an 8.4\,m primary mirror, using the LBT Utility Camera in the Infrared (LUCI) instrument (programme ID: UV-2022A-004, PI: J.\,C. Tan). Adaptive Optics (AO) assisted mode was used with the Single conjugated adaptive Optics Upgrade for LBT \citep[SOUL][]{pinna2016}. The N30 camera with a pixel scale of 0\farcs015 and field of view (FoV) of $30\arcsec\times30\arcsec$ was used. The filters K$_{\rm s}$, H$_2$, and Br$\gamma$ which are centred at the wavelengths  2.163,  2.124,  2.170\,$\mu{\rm m}$, respectively, were employed.

Images were centred around S255IR NIRS3 (with central coordinates of the image RA(J2000)=06:12:54.355, Dec(J2000)=+17:59:23.748). The AO guide star used (2MASSJ06125505+1759289 RA(J2000)=06:12:55.049, Dec(J2000)=+17:59:28.896, R=14.8\,mag) is located at 14\arcsec from NIRS3.
% The AO guide star used was 2MASSJ06125505+1759289 (RA(J2000)=06:12:55.049, Dec(J2000)=+17:59:28.896) from which the telescope wave sensor detected R=14.8\,mag and it was at a distance of 14\arcsec\:from the central coordinates of the image (RA(J2000)=06:12:54.355, Dec(J2000)=+17:59:23.748), centred around the location of NIRS3.
% It should be noted that the AO guide star used in our observations is the same star that \citet{caratti2017nature} used in their observations.
The Strehl ratio was 0.2 and the final Full Width Half Maximum (FWHM) at the position of the AO guide star was $\sim0\farcs06$ for all three filters (derived by fitting Moffat profiles, see Appendix\,\ref{sect:appendix_soul} for more details on SOUL performance), namely close to the LBT diffraction limit.

The data were reduced and flux calibrated with custom python scripts using the python packages ccdproc \citep{ccdproc}, astropy \citep{astropy2018} and photutils \citep{photutils2020}. The flux calibration was performed by matching field stars with the UKIRT Infrared Deep Sky Survey (UKIDSS) point source catalogue, achieving an accuracy of $\sim0.06$\,mag. The data were astrometrically corrected by matching stars to the Gaia DR3 catalogue retrieved with astroquery. We were able to match 7 Gaia stars in our FoV and have a final residual for the astrometry of 0\farcs034.

We also present Atacama Large Millimeter/submillimeter Array (ALMA) band 3 observations taken on the 3rd of September 2021 with baselines ranging from 122\,m to 1619\,m (programme ID: 2018.1.00864.S, PI: R. Cesaroni) and having a synthesised beam of 0\farcs87. In this work, we only present the band 3 continuum image and defer further analysis of this dataset to a future paper (Cesaroni et al. in prep.).

\section{Results and Discussion}\label{sect:results}

Here, we present the main results of our imaging on
% the high-mass star-forming region S255IR focusing on 
the outbursting source NIRS3. The left panel of Figure\,\ref{fig:continuum_image} shows our AO-assisted K$_{\rm s}$ continuum image for the central $\sim30\arcsec\times30\arcsec$ of S255IR. The sub-millimetre sources are labelled as reported by \citet{zinchenko2020}, where the two main sources are SMA1, \citep[called S255-2c in][]{snell1986} or NIRS3 in this work; and SMA3 or NIRS1 in this work. The two bright nebulous regions to the east and to the west of NIRS3 were reported by \citet{tamura1991} as the red and blue lobes of a large-scale outflow, respectively. The right panel of Figure\,\ref{fig:continuum_image} shows a red-green-blue (RGB) image using the filters H$_2$-Br$\gamma$-K$_{\rm s}$, respectively. Centred at NIRS3, there is a chain of H$_2$ knots towards the northeast (NE) and southwest (SW) directions, as revealed in the red channel, that could be associated with previous ejection episodes (labelled as `NIRS3 H$_2$ knots'). The NE jet lobe position angle (PA) as traced by these knots is $\sim30^\circ{}$, whereas the PA for the SW lobe is $\sim230^\circ{}$, which is roughly in the opposite direction of that of the NE lobe, consistent with previous studies \citep{tamura1991}. Towards the NE, the knots observed extend up to $8\farcs5$ which corresponds to a projected distance of $\sim15\,000$\,au from NIRS3. In the opposite direction, SW, the knots extend up to $\sim5\arcsec$ ($\sim8900$\,au projected) from NIRS3.

From our AO imaging data, we can estimate the ejection time scale of these H$_2$ knots. It has been observed that H$_2$ knots driven by HMYSOs can have velocities exceeding $100\,\mathrm{km\,s^{-1}}$ \citep[see, e.g.,][]{fedriani2018,massi2023}. This would imply that the dynamic age of the NIRS3 H$_2$ knots is $\sim400-700$ years. It should be noted that our images have a FoV of $\sim30\arcsec\times30\arcsec$, therefore it is possible that we are not observing the leading outer bow shocks of the flow. In any case, this H$_2$ emission seems to be tracing shocked jet emission consistent with what was found by \citet{howard1997} and \citet{miralles1997}. We also identify a number of extra knots towards the east of the main S255IR complex that seems not to be associated with NIRS3 (labelled `Other H$_2$ knots'). In the following, we will focus on the much smaller 3500\,au$^2$ area around NIRS3.

\subsection{Episodic Ejection as a Response to Episodic Accretion}

% Ks LBT image combined with VLA and ALMA
\begin{figure}[!htb]
    \includegraphics[width=0.49\textwidth]{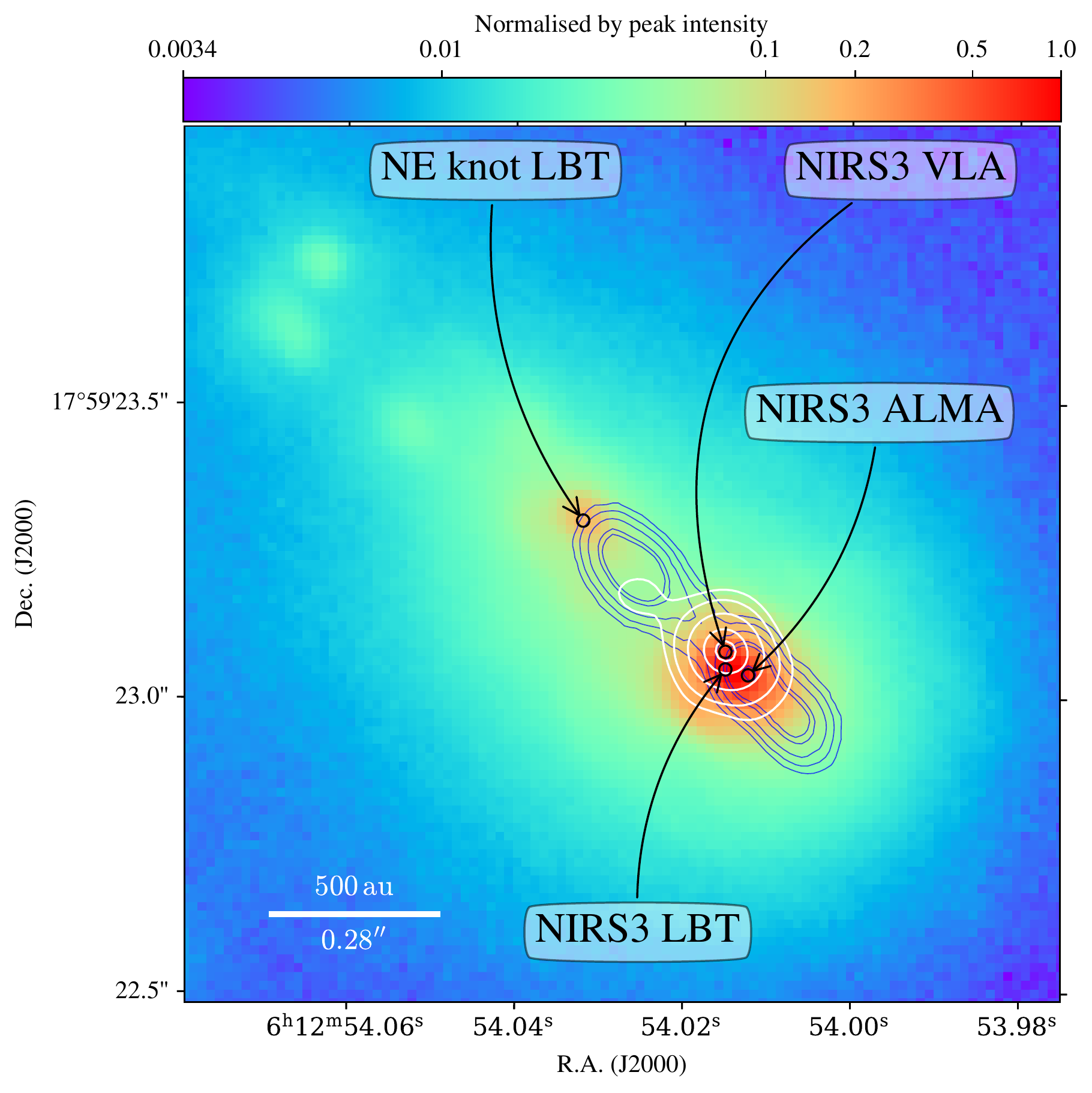}
  \caption{AO-assisted K$_{\rm s}$ band image zoom-in into the central region of S255IR featuring NIRS3. The image is normalised to the peak intensity from NIRS3 (labelled NIRS3 LBT) and scaled logarithmically as shown in the colorbar on top of the panel. Blue contours represent ALMA band 3 continuum emission with contour levels [0.5, 1.0, 2.0, 4.0, 5.0, 10.0] $\times10^{-3}$\,Jy/beam (Cesaroni et al. in prep.). White contours show VLA 1.3\,cm emission with contour levels [0.8,2.0,5.0,10.0,13.2] $\times10^{-3}$\,Jy/beam from \citet{cesaroni2018}. Black circles represent the peak position for NIRS3 and NE knot for LBT, ALMA, and VLA as labelled.}
     \label{fig:S255_multiband}
\end{figure}

% \textcolor{red}{RF: Could this be the closest detection of a jet knot in a massive YSO in the NIR (both in continuum and in H2)? is it worth mentioning?}

In June 2015, NIRS3 in the S255IR region underwent an accretion outburst \citep{fujisawa2015,caratti2017nature}. Using the VLA, \citet{cesaroni2018} reported a radio jet flare arising from this  accretion outburst. They also present evidence of a possible new radio jet knot at $\sim0\farcs17$ ($\sim300$\,au) from the source in December 2016. In Figure\,\ref{fig:S255_multiband} we show our February 2022 K$_{\rm s}$ band image of the region around NIRS3 (see also Figure\,\ref{fig:H2_and_BrG_images} for H$_2$ and Br$\gamma$ images). In this Figure, we also show as white contours the VLA observations (December 2016) presented in Figure 4 from \citet{cesaroni2018} showing the new knot (NE from the source). We also present unpublished ALMA band 3 observations (September 2021) as blue contours that trace the extended free-free emission from the protostellar jet while the peak emission to the SW pinpoints the location of the massive protostar (Cesaroni et al. in prep.). The intensity peaks from the radio and NIR images are labelled as `NIRS3 ALMA', `NIRS3 VLA', and `NIRS3 LBT', respectively. There is a shift of $\sim3$ pixels ($\sim0\farcs045$) between the positions of the LBT and ALMA peaks, which is close to our astrometric accuracy (see Sect.\,\ref{sect:observations}), and therefore they may well coincide, although in most cases this does not happen, depending on the physical properties, including the geometry of the system. The NIR peak at the position of NIRS3 is interpreted as reflected light tracing the combined emission of accretion activity and ejection of material very close to the star.
% However, radio and NIR emission peaks often do not coincide, depending on the physical properties, including the geometry of the system. Nonetheless, the NIR peak can be interpreted as a combination of the emission of accretion activity, ejection of material very close to the star, and reflected light. The latter is the most likely explanation for the NIR extended emission around NIRS3.

Towards the north east of NIRS3, there is another point-like source labelled `NE knot LBT' (see Figure\,\ref{fig:continuum_image_non-annotated} for a version without annotations or contours). We interpret this NIR emission as a jet knot and suggest that it is the same jet knot observed with the VLA \citep{cesaroni2018}, 6 years earlier from the LBT observations. We measure a projected distance between the NIR NE knot and NIRS3 of $\sim0\farcs35$, which corresponds to $\sim634$\,au at the distance of the source. Assuming that this knot was ejected shortly after the accretion outburst (i.e. $\Delta t\sim2435$\,days, which is the difference between the outburst onset and our observations), the mean tangential velocity is $\sim450\pm50\,\mathrm{km\,s^{-1}}$ (considering astrometric errors). The inclination of the accretion disk of NIRS3 is $\sim80^\circ{}$ \citep{boley2013} being almost edge-on. This allows us to calculate the total velocity of the jet, assuming that disk and jet are perpendicular, which is $\sim458\,\mathrm{km\,s^{-1}}$. This velocity is consistent with previously measured jet knots ejected by massive protostars \citep[e.g.][]{fedriani2019,massi2023}. This interpretation is supported by the fact that the other jet tracers in the radio regime align with the NIR NE knot. There seems to be a temporal evolution of the jet knot between the radio and NIR observations. In particular, the ALMA emission, which was observed between the VLA and LBT observations, is roughly in the middle of the radio and NIR knot. As HMYSOs are highly embedded, it is rare to observe NIR and radio jet emission spatially coincident near the source, with only another example in the literature reported by \citet{fedriani2019}. We also start to see the counter jet towards the SW of NIRS3 in the form of 3\,mm continuum elongated emission. There is no NIR jet knot counterpart at this location possibly because the extinction is higher and obscures the infrared emission.

% Continuum-subtracted images
\begin{figure*}[!htb]
        \includegraphics[width=0.48\textwidth]{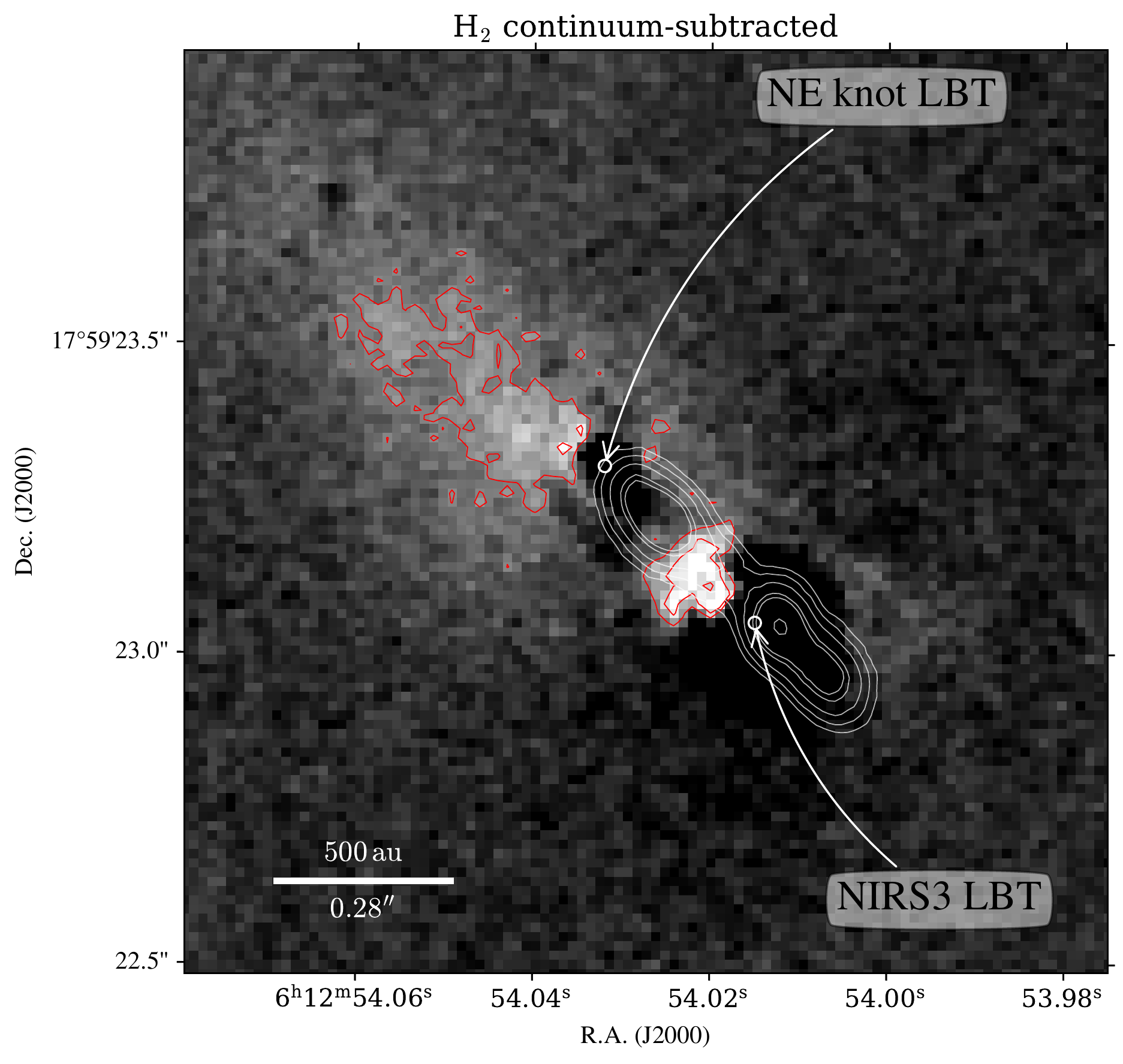}
        \includegraphics[width=0.48\textwidth]{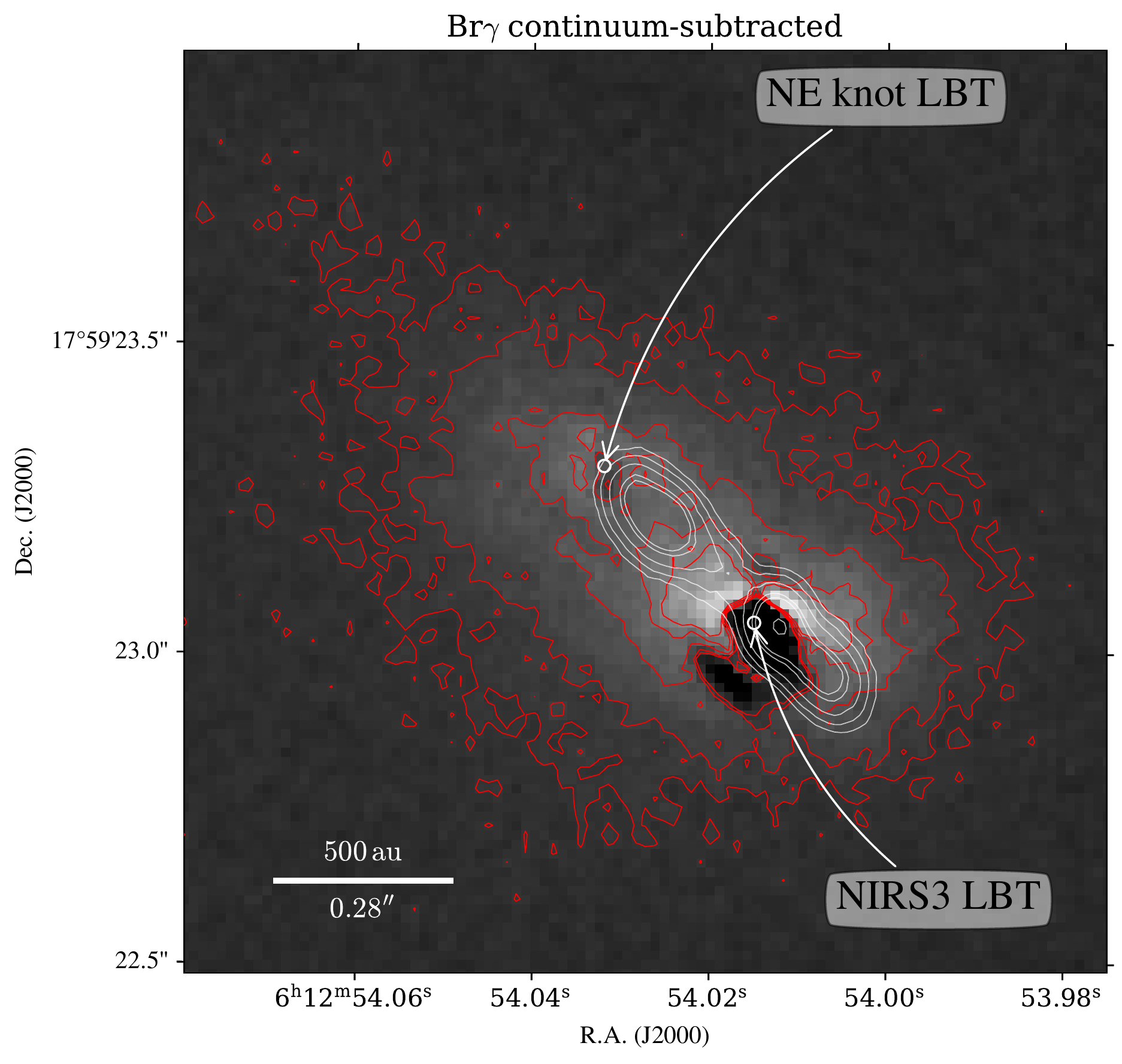}

  \caption{Continuum-subtracted images for the surrounding region of the outbursting source NIRS3. Left panel: H$_2$ continuum-subtracted. Right panel: Br$\gamma$ continuum-subtracted. Red contours represent emission at 3, 5, and 10$\times$RMS, where RMS is the root mean square for each image. White contours are ALMA band 3 continuum emission and contours levels are [0.5, 1.0, 2.0, 4.0, 5.0, 10.0] $\times10^{-3}$\,Jy/beam.}
     \label{fig:subtracted_images}
\end{figure*}

Figure\,\ref{fig:subtracted_images} shows the continuum-subtracted images for H$_2$ and Br$\gamma$, left and right panels respectively. To obtain the continuum-subtracted images we measure the flux of nearby isolated stars in both the broad-band (K$_{\rm s}$) and narrow-band (H$_2$ and Br$\gamma$) filters. We then calculate the ratio of the integrated fluxes between the broad- and narrow-band filters. Finally, the multiplicative factor to be applied to the images is calculated by obtaining the median value of the ratio of the fluxes. This method assumes approximately flat continua between the broad and narrow filters, given its wavelength proximity. The greyscale images show the result of the continuum-subtracted method and the red contours represent 3, 5, and 10 times the root mean square (RMS) of each image. It should be noted that using this technique is often the case that the strong NIR continuum of the main massive protostar becomes over-subtracted as evinced by the black dips in the images. Nonetheless, the H$_2$ continuum-subtracted emission clearly shows an elongated structure towards the NE of NIRS3. This emission is consistent with jet shocked activity as discussed by \citet{howard1997} and \citet{miralles1997}. It should be noted that the H$_2$ emission coincides with the VLA radio jet, which is ionised emission from shocks. This supports our interpretation of collisional origin for the H$_2$. The Br$\gamma$ continuum-subtracted image shows a different morphology. It shows a less elongated structure and more diffuse emission than the H$_2$. We interpret this emission as scattered light in the outflow cavity walls from the central system. This scattered light would come from a combination of the accretion activity occurring very close to the central engine and the UV radiation coming directly from the massive protostar. Even though \citet{miralles1997} and \citet{wang2011} suggested excitation by collisions for the H$_2$ emission, their observations were limited by low spectral resolution ($R\sim700$ and $R\sim1500$, respectively). High spectral resolution observations are needed to measure the kinematic and dynamic properties of these knots and establish their nature.

\subsection{Thirty Year Photometric Variability in NIRS3}\label{sect:photometry}

Variability in low-mass YSOs has been observed in many systems \citep[see, e.g.,][]{herbig1977,hartmann1996,macfarlane2019,zakri2022} and simulated \citep{zhu2009,stamatellos2011,stamatellos2012}, see \citet{fischer2022} for a recent review. This variability has many origins, one of which is accretion variability giving rise to outbursts. However, accretion variability in high-mass protostars is less explored with only a handful of observed examples \citep{caratti2017nature,hunter2017,stecklum2021,chen2021}. 3D simulations have recently addressed this episodic accretion scenario in HMYSOs \citep{elbakyan2023}. The simulations suggest that high-density post-collapse clumps crossing the inner computational boundary could give rise to observable outbursts.

\citet{caratti2017nature} reported the accretion outburst of NIRS3 by comparing a pre-outburst image from UKIDSS with a (post)outburst image from PANIC \citep{stecklum2016}, dated 8th December 2009 and 28th November 2015, respectively. The authors measured a large increase in both the H and K bands by $\Delta {\rm H}\sim3.5$\,mag $\Delta {\rm K}\sim2.5$\,mag. There was a simultaneous flaring of the 6.7\,GHz methanol masers \citep{fujisawa2015}. Soon after, \citet{uchiyama2020} performed a NIR follow-up of NIRS3 and measured the photometry from November 2015
% when they measured K=$8.683\pm0.023$\,mag
to March 2019,
% when they measured K=$11.747\pm0.050$\,mag.
measuring the maximum brightness in December 2015 with K=$8.03\pm0.05$\,mag. The authors discuss that in the $\sim4.5$\,yr baseline of observations, they only detect one episodic event. However, we have gathered
photometry from the late 1980s to the early 2000s and we can speculate that another episodic event took place in the late 1980s \citet{tamura1991} measured a K$=10.1$\,mag in NIRS3 in May 1989, although the authors state the poor accuracy of this photometric point that may be caused by contamination of nearby sources. Then, \citet{hodapp1994} observed this region in February 1991 and we perform aperture photometry on their images obtaining K$=11.49\pm0.03$\,mag. \citet{howard1997} and \citet{miralles1997} reported K$=11.08$\,mag in February 1994 and K$=11.31$\,mag in February 1996, respectively, for NIRS3. Later, in February 1999, \citet{bik2004thesis} reported K$=12.56$\,mag, which seems to indicate the minimum of this alleged previous outburst. Only a year later, in 2000, \citet{itoh2001} and \citet{ojha2011} report K$=12.01$\,mag and  K$=11.82$\,mag, respectively, indicating that the previous photometric point was likely a minimum. Although the photometric dataset is not homogeneous and the observations were taken with different instruments and conditions, this dataset hints that NIRS3 is highly variable.

\begin{figure}[!htb]
    \includegraphics[width=0.49\textwidth]{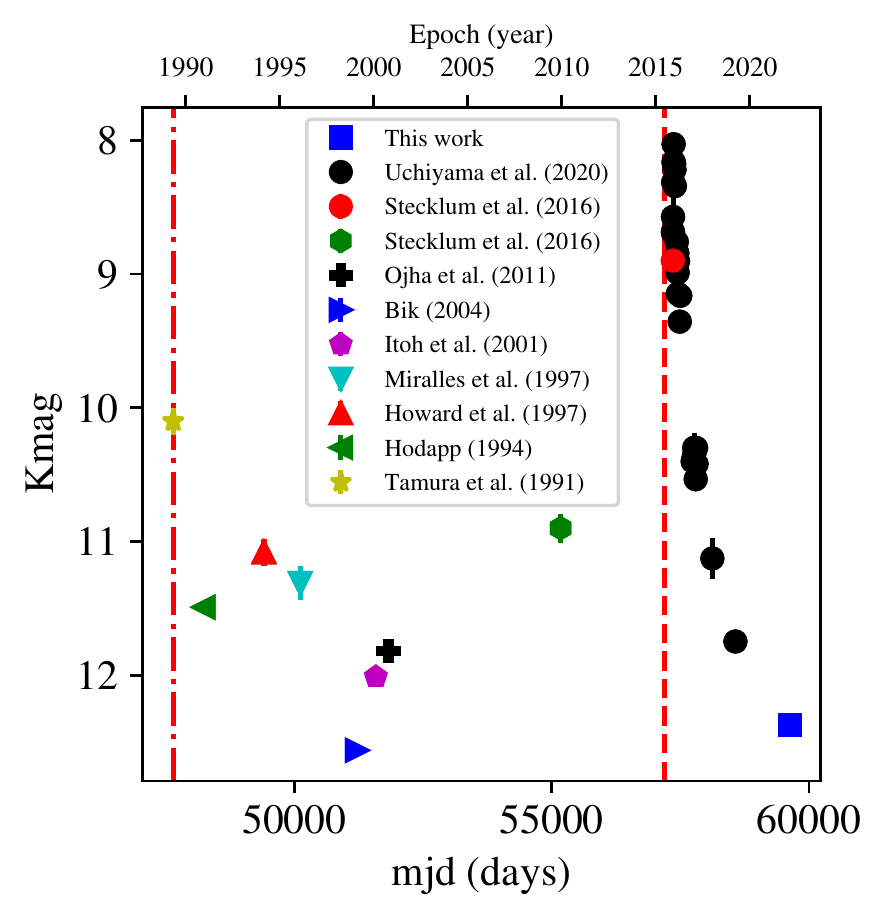}
  \caption{Photometry for S255IR NIRS3 source. The lower x-axis is in units of modified Julian date (mjd) in days and the upper x-axis gives the date. Errors are also plotted, but some of them are smaller than the symbol used. The red dashed-dotted line represents the alleged outburst that occurred in the late 1980s, whereas the red dashed line marks the confirmed outburst registered in June 2015.}
     \label{fig:photometry_S255}
\end{figure}

In our AO-assisted images, we have performed photometry at the position of NIRS3 (see Table\,\ref{tab:soul_performance} for exact coordinates used). Here, we report two values for our 2022 epoch, one with an aperture radius of 1\arcsec, in order to compare with previous studies, and the other with radius $\sim2\times\mathrm{FWHM}\sim0\farcs15$ to avoid scattered light from the surrounding medium (see Figure \ref{fig:aperture_phot_S255}). In the first case, we measure $\mathrm{K_{1\arcsec}}=12.37\pm0.06$\,mag whereas in the second $\mathrm{K_{0\farcs15}}=13.48\pm0.06$\,mag. NIRS3 seems to be in its dimmest state in the historic series. The photometry was performed using the tool SedFluxer from sedcreator \citep{fedriani2023}, see Appendix\,\ref{sect:sedcreator} for more details. Figure\,\ref{fig:photometry_S255} shows the K band photometry for NIRS3 from 1989 to 2022. The $\sim30$ years of photometry for NIRS3 suggests that in addition to the 2015 outburst, there may have been another one in the late 1980s (indicated as red dashed and dashed-dotted lines). The amplitude of the first episode is $\Delta K\gtrsim-2.5$\,mag. The quiescent phase then lasted about $10$ years as the UKIDSS photometric point starts to raise with respect to that of \citet{itoh2001}. After that, NIRS3 underwent the confirmed outburst \citep{fujisawa2015,stecklum2016,caratti2017nature} with an amplitude of $\Delta K\sim3.37$\,mag, considering the UKIDSS photometry and the maximum of \citet{uchiyama2020}. After a further 7 years, the NIR K magnitude dropped by $\Delta K\sim-4.34$\,mag (or $\Delta K\sim-5.45$\,mag if we consider $\mathrm{K_{0\farcs15}}$) from the maximum in \citet{uchiyama2020}. To the best of our knowledge, this is the largest amplitude episode ever registered for NIRS3 and for any high-mass protostar in the K band (assuming that the visual extinction towards the source did not dramatically change). Interestingly, the time span between the bright and dim states would be similar in both episodic events, i.e., $\sim6-10$\,yr. Table\,\ref{tab:photometry} summarises the photometric points discussed above and presented in Figure\,\ref{fig:photometry_S255}.

Although there are no methanol maser observations at the time of the alleged first outburst (late 1980s), we note that the NIR photometric variability seen in the 1990s is consistent with the maser variability reported in the same period. \citet{caswell1995} reported a 10\% variability in the methanol maser between 1992 and 1993 and an increase of a factor of 10 of the feature at $1.8\,\mathrm{km\,s^{-1}}$. This is a remarkable change in methanol masers activity as these masers do not usually change by a factor of 10. For context, \citet{szymczak2018} performed monitoring of methanol masers in 139 star-forming sites and reported that 80\% of the sources display variability greater than 10\%. They discuss the scenario of short-lived outbursts for those sites with a relative amplitude greater than 2. Moreover, as part of the long-term program, \citet{szymczak2018S255} present methanol maser observations for S255IR NIRS3, including previous works by \citet{menten1991}, \citet{caswell1995}, \citet{szymczak2000}, \citet{goedhart2004}, and \citet{szymczak2012}. \citet{szymczak2018S255} show the available flux measurements of the methanol maser from 1991 to 2017 (see their Fig. 2), a period that does not cover the alleged outburst in the
late 1980s. However, it should be noted that changes in the pumping radiation due to episodic accretion is only one of the possible reasons for maser flux variability. Therefore, it seems possible that from the late 1980s to the early 2000s, there have been fluctuations in the accretion rate onto the protostar as suggested by the NIR photometric and methanol maser variability. The earlier NIR observations by \citet{strom1976} and \citet{evans1977} do not report photometric measurements for NIRS3, thus cannot be used to confirm the existence of an outburst.

\citet{burns2016} propose that at least three episodes of ejections have taken place in NIRS3 (which they refer to as SMA1) in the last few thousand years. The signatures of one of these ejection episodes was observed by \citet{wang2010} and \citet{zinchenko2015} in the form of [FeII] and HCO$^+$ jet knots, respectively, with a dynamic age of $\sim1000$\,yr. The observed NIRS3 H$_2$ knots indicated by the white arrows in the left panel of Figure\,\ref{fig:continuum_image} would be part of this episode as their dynamic age is similar (400-700\,yr, see Sect.\,\ref{sect:results}). The most recent ejection episode is then our radio/NIR `NE knot LBT' with just 7 years of dynamic age (see Fig.\,\ref{fig:continuum_image} and \ref{fig:subtracted_images}).

\section{Conclusions}\label{sect:conclusions}

We have observed the high-mass star-forming region S255IR using the Large Binocular Telescope with the LUCI instrument in AO-assisted imaging mode achieving an angular resolution of $\sim0\farcs06$. S255IR harbours the outbursting source NIRS3 which underwent an accretion outburst in 2015 and has faded since then. The photometry measured in this work yields the faintest minimum to date for NIRS3 with $\mathrm{K_{0\farcs15}}=13.48$\,mag, indicating that is in a quiescent state.

We observed a new knot NE of NIRS3 that we interpret as a jet knot ejected during the latest accretion outburst. We measured a  tangential velocity of $450\pm50\,\mathrm{km\,s^{-1}}$ for this knot. This emission is consistent with previous radio observations made with the VLA and ALMA tracing the outflowing material. In the H$_2$ continuum-subtracted image, we identify extended elongated emission consistent with jet-shocked material. On the other hand, in the Br$\gamma$ continuum-subtracted image, the extended emission is more likely to be scattered light from a combination of the accretion activity and the UV radiation from the massive protostar NIRS3.

We speculate on the possibility that NIRS3 has undergone two outbursts in the last 30 years. NIRS3 would then be the first massive protostar with at least two outbursts observed in the NIR. Although we cannot firmly confirm the first burst, the high photometric variability (along with other accretion-burst tracers) suggests that NIRS3 is in a very active accretion stage prone to more accretion bursts.

\begin{acknowledgements}
    The authors acknowledge fruitful discussions with G. MacLeod and R. Burns. R.F. acknowledges support from the grants Juan de la Cierva FJC2021-046802-I, PID2020-114461GB-I00 and CEX2021-001131-S funded by MCIN/AEI/ 10.13039/501100011033 and by ``European Union NextGenerationEU/PRTR'' and grant P20-00880 from the Consejer\'ia de Transformaci\'on Econ\'omica, Industria, Conocimiento y Universidades of the Junta de Andaluc\'ia. R.F. also acknowledges funding from the European Union’s Horizon 2020 research and innovation programme under the Marie Sklodowska-Curie grant agreement No 101032092.  A.C.G. has been supported by PRIN-INAF-MAIN-STREAM 2017 ``Protoplanetary disks seen through the eyes of new-generation instruments'' and by PRIN-INAF 2019 ``Spectroscopically tracing the disk dispersal evolution (STRADE)''. J.C.T. acknowledges support from ERC grant MSTAR, VR grant 2017-04522, and NSF grant 1910675. G.C. acknowledges support from the Swedish Research Council (VR Grant; Project: 2021-05589).
\end{acknowledgements}

\bibliographystyle{aa}
\bibliography{phd_bibliography.bib}

\begin{thebibliography}{56}
\expandafter\ifx\csname natexlab\endcsname\relax\def\natexlab#1{#1}\fi

\bibitem[{{Astropy Collaboration} {et~al.}(2018){Astropy Collaboration},
  {Price-Whelan}, {Sip{\H o}cz}, {G{\"u}nther}, {Lim}, {Crawford}, {Conseil},
  {Shupe}, {Craig}, {Dencheva}, {Ginsburg}, {VanderPlas}, {Bradley},
  {P{\'e}rez-Su{\'a}rez}, {de Val-Borro}, {Aldcroft}, {Cruz}, {Robitaille},
  {Tollerud}, {Ardelean}, {Babej}, {Bach}, {Bachetti}, {Bakanov}, {Bamford},
  {Barentsen}, {Barmby}, {Baumbach}, {Berry}, {Biscani}, {Boquien}, {Bostroem},
  {Bouma}, {Brammer}, {Bray}, {Breytenbach}, {Buddelmeijer}, {Burke},
  {Calderone}, {Cano Rodr{\'{\i}}guez}, {Cara}, {Cardoso}, {Cheedella},
  {Copin}, {Corrales}, {Crichton}, {D'Avella}, {Deil}, {Depagne}, {Dietrich},
  {Donath}, {Droettboom}, {Earl}, {Erben}, {Fabbro}, {Ferreira}, {Finethy},
  {Fox}, {Garrison}, {Gibbons}, {Goldstein}, {Gommers}, {Greco}, {Greenfield},
  {Groener}, {Grollier}, {Hagen}, {Hirst}, {Homeier}, {Horton}, {Hosseinzadeh},
  {Hu}, {Hunkeler}, {Ivezi{\'c}}, {Jain}, {Jenness}, {Kanarek}, {Kendrew},
  {Kern}, {Kerzendorf}, {Khvalko}, {King}, {Kirkby}, {Kulkarni}, {Kumar},
  {Lee}, {Lenz}, {Littlefair}, {Ma}, {Macleod}, {Mastropietro}, {McCully},
  {Montagnac}, {Morris}, {Mueller}, {Mumford}, {Muna}, {Murphy}, {Nelson},
  {Nguyen}, {Ninan}, {N{\"o}the}, {Ogaz}, {Oh}, {Parejko}, {Parley}, {Pascual},
  {Patil}, {Patil}, {Plunkett}, {Prochaska}, {Rastogi}, {Reddy Janga},
  {Sabater}, {Sakurikar}, {Seifert}, {Sherbert}, {Sherwood-Taylor}, {Shih},
  {Sick}, {Silbiger}, {Singanamalla}, {Singer}, {Sladen}, {Sooley},
  {Sornarajah}, {Streicher}, {Teuben}, {Thomas}, {Tremblay}, {Turner},
  {Terr{\'o}n}, {van Kerkwijk}, {de la Vega}, {Watkins}, {Weaver}, {Whitmore},
  {Woillez}, {Zabalza}, \& {Astropy Contributors}}]{astropy2018}
{Astropy Collaboration}, {Price-Whelan}, A.~M., {Sip{\H o}cz}, B.~M., {et~al.}
  2018, \aj, 156, 123

\bibitem[{{Bally}(2016)}]{bally2016}
{Bally}, J. 2016, \araa, 54, 491

\bibitem[{{Beltr{\'a}n} \& {de Wit}(2016)}]{beltran2016}
{Beltr{\'a}n}, M.~T. \& {de Wit}, W.~J. 2016, \aapr, 24, 6

\bibitem[{Bik(2004)}]{bik2004thesis}
Bik, A. 2004, Ph. D. Thesis

\bibitem[{{Boley} {et~al.}(2013){Boley}, {Linz}, {van Boekel}, {Henning},
  {Feldt}, {Kaper}, {Leinert}, {M{\"u}ller}, {Pascucci}, {Robberto},
  {Stecklum}, {Waters}, \& {Zinnecker}}]{boley2013}
{Boley}, P.~A., {Linz}, H., {van Boekel}, R., {et~al.} 2013, \aap, 558, A24

\bibitem[{Bradley {et~al.}(2020)Bradley, Sip{\H o}cz, Robitaille, Tollerud,
  Vin{\'{\i}}cius, Deil, Barbary, Wilson, Busko, G{\"u}nther, Cara, Conseil,
  Bostroem, Droettboom, Bray, Bratholm, Lim, Barentsen, Craig, Pascual, Perren,
  Greco, Donath, de~Val-Borro, Kerzendorf, Bach, Weaver, D'Eugenio, Souchereau,
  \& Ferreira}]{photutils2020}
Bradley, L., Sip{\H o}cz, B., Robitaille, T., {et~al.} 2020, astropy/photutils:
  1.0.0

\bibitem[{{Burns} {et~al.}(2016){Burns}, {Handa}, {Nagayama}, {Sunada}, \&
  {Omodaka}}]{burns2016}
{Burns}, R.~A., {Handa}, T., {Nagayama}, T., {Sunada}, K., \& {Omodaka}, T.
  2016, \mnras, 460, 283

\bibitem[{{Cabrit}(2007)}]{cabrit2007}
{Cabrit}, S. 2007, in IAU Symposium, Vol. 243, Star-Disk Interaction in Young
  Stars, ed. J.~{Bouvier} \& I.~{Appenzeller}, 203--214

\bibitem[{{Caratti o Garatti} {et~al.}(2017){Caratti o Garatti}, {Stecklum},
  {Garcia Lopez}, {Eisl{\"o}ffel}, {Ray}, {Sanna}, {Cesaroni}, {Walmsley},
  {Oudmaijer}, {de Wit}, {Moscadelli}, {Greiner}, {Krabbe}, {Fischer}, {Klein},
  \& {Iba{\~n}ez}}]{caratti2017nature}
{Caratti o Garatti}, A., {Stecklum}, B., {Garcia Lopez}, R., {et~al.} 2017,
  Nature Physics, 13, 276

\bibitem[{{Caswell} {et~al.}(1995){Caswell}, {Vaile}, \&
  {Ellingsen}}]{caswell1995}
{Caswell}, J.~L., {Vaile}, R.~A., \& {Ellingsen}, S.~P. 1995, \pasa, 12, 37

\bibitem[{{Cesaroni} {et~al.}(2018){Cesaroni}, {Moscadelli}, {Neri}, {Sanna},
  {Caratti o Garatti}, {Eisloffel}, {Stecklum}, {Ray}, \&
  {Walmsley}}]{cesaroni2018}
{Cesaroni}, R., {Moscadelli}, L., {Neri}, R., {et~al.} 2018, \aap, 612, A103

\bibitem[{{Chen} {et~al.}(2021){Chen}, {Sun}, {Chini}, {Haas}, {Jiang}, \&
  {Chen}}]{chen2021}
{Chen}, Z., {Sun}, W., {Chini}, R., {et~al.} 2021, \apj, 922, 90

\bibitem[{Craig {et~al.}(2022)Craig, Crawford, Seifert, Robitaille, Sipőcz,
  Walawender, Crawford, Vinícius, Ninan, Droettboom, Youn, Gondhalekar,
  Tollerud, Bowers, Lim, Bray, Bach, stottsco, Janga, walkerna22, Günther,
  Rol, A., Bradley, Price-Whelan, Deil, Ryon, \& Lee}]{ccdproc}
Craig, M., Crawford, S., Seifert, M., {et~al.} 2022, {astropy/ccdproc: 2.3.1 --
  fixes astropy 5.1 compatibility}

\bibitem[{{Elbakyan} {et~al.}(2023){Elbakyan}, {Nayakshin}, {Meyer}, \&
  {Vorobyov}}]{elbakyan2023}
{Elbakyan}, V.~G., {Nayakshin}, S., {Meyer}, D. M.~A., \& {Vorobyov}, E.~I.
  2023, \mnras, 518, 791

\bibitem[{{Evans} {et~al.}(1977){Evans}, {Blair}, \& {Beckwith}}]{evans1977}
{Evans}, N.~J., I., {Blair}, G.~N., \& {Beckwith}, S. 1977, \apj, 217, 448

\bibitem[{{Fedriani} {et~al.}(2018){Fedriani}, {Caratti o Garatti}, {Coffey},
  {Lopez}, {Kraus}, {Weigelt}, {Stecklum}, {Ray}, \& {Walmsley}}]{fedriani2018}
{Fedriani}, R., {Caratti o Garatti}, A., {Coffey}, D., {et~al.} 2018, \aap,
  616, A126

\bibitem[{{Fedriani} {et~al.}(2019){Fedriani}, {Garatti}, {Purser}, {Sanna},
  {Tan}, {Garcia-Lopez}, {Ray}, {Coffey}, {Stecklum}, \&
  {Hoare}}]{fedriani2019}
{Fedriani}, R., {Garatti}, A. C.~o., {Purser}, S.~J.~D., {et~al.} 2019, Nature
  Communications, 10, 3630

\bibitem[{{Fedriani} {et~al.}(2023){Fedriani}, {Tan}, {Telkamp}, {Zhang},
  {Yang}, {Liu}, {De Buizer}, {Law}, {Beltran}, {Rosero}, {Tanaka},
  {Cosentino}, {Gorai}, {Farias}, {Staff}, \& {Whitney}}]{fedriani2023}
{Fedriani}, R., {Tan}, J.~C., {Telkamp}, Z., {et~al.} 2023, \apj, 942, 7

\bibitem[{{Fischer} {et~al.}(2022){Fischer}, {Hillenbrand}, {Herczeg},
  {Johnstone}, {K{\'o}sp{\'a}l}, \& {Dunham}}]{fischer2022}
{Fischer}, W.~J., {Hillenbrand}, L.~A., {Herczeg}, G.~J., {et~al.} 2022, arXiv
  e-prints, arXiv:2203.11257

\bibitem[{{Fujisawa} {et~al.}(2015){Fujisawa}, {Yonekura}, {Sugiyama},
  {Horiuchi}, {Hayashi}, {Hachisuka}, {Matsumoto}, \& {Niinuma}}]{fujisawa2015}
{Fujisawa}, K., {Yonekura}, Y., {Sugiyama}, K., {et~al.} 2015, The Astronomer's
  Telegram, 8286, 1

\bibitem[{{Goedhart} {et~al.}(2004){Goedhart}, {Gaylard}, \& {van der
  Walt}}]{goedhart2004}
{Goedhart}, S., {Gaylard}, M.~J., \& {van der Walt}, D.~J. 2004, \mnras, 355,
  553

\bibitem[{{Hartmann} \& {Kenyon}(1996)}]{hartmann1996}
{Hartmann}, L. \& {Kenyon}, S.~J. 1996, \araa, 34, 207

\bibitem[{{Herbig}(1977)}]{herbig1977}
{Herbig}, G.~H. 1977, \apj, 217, 693

\bibitem[{{Hodapp}(1994)}]{hodapp1994}
{Hodapp}, K.-W. 1994, \apjs, 94, 615

\bibitem[{{Howard} {et~al.}(1997){Howard}, {Pipher}, \& {Forrest}}]{howard1997}
{Howard}, E.~M., {Pipher}, J.~L., \& {Forrest}, W.~J. 1997, \apj, 481, 327

\bibitem[{{Hunter} {et~al.}(2017){Hunter}, {Brogan}, {MacLeod}, {Cyganowski},
  {Chandler}, {Chibueze}, {Friesen}, {Indebetouw}, {Thesner}, \&
  {Young}}]{hunter2017}
{Hunter}, T.~R., {Brogan}, C.~L., {MacLeod}, G., {et~al.} 2017, \apjl, 837, L29

\bibitem[{{Itoh} {et~al.}(2001){Itoh}, {Tamura}, {Suto}, {Hayashi}, {Murakawa},
  {Oasa}, {Nakajima}, {Kaifu}, {Kosugi}, {Usuda}, \& {Doi}}]{itoh2001}
{Itoh}, Y., {Tamura}, M., {Suto}, H., {et~al.} 2001, \pasj, 53, 495

\bibitem[{{Lo} {et~al.}(1975){Lo}, {Burke}, \& {Haschick}}]{lo1975}
{Lo}, K.~Y., {Burke}, B.~F., \& {Haschick}, A.~D. 1975, \apj, 202, 81

\bibitem[{{MacFarlane} {et~al.}(2019){MacFarlane}, {Stamatellos}, {Johnstone},
  {Herczeg}, {Baek}, {Chen}, {Kang}, \& {Lee}}]{macfarlane2019}
{MacFarlane}, B., {Stamatellos}, D., {Johnstone}, D., {et~al.} 2019, \mnras,
  487, 5106

\bibitem[{{Massi} {et~al.}(2023){Massi}, {Caratti o Garatti, A.}, {Cesaroni,
  R.}, {Sridharan, T. K.}, {Ghose, E.}, {Pinna, E.}, {Beltr\'an, M. T.},
  {Leurini, S.}, {Moscadelli, L.}, {Sanna, A.}, {Agapito, G.}, {Briguglio, R.},
  {Christou, J.}, {Esposito, S.}, {Mazzoni, T.}, {Miller, D.}, {Plantet, C.},
  {Power, J.}, {Puglisi, A.}, {Rossi, F.}, {Rothberg, B.}, {Taylor, G.}, \&
  {Veillet, C.}}]{massi2023}
{Massi}, F., {Caratti o Garatti, A.}, {Cesaroni, R.}, {et~al.} 2023, A\&A, 672,
  A113

\bibitem[{{Menten}(1991)}]{menten1991}
{Menten}, K.~M. 1991, \apjl, 380, L75

\bibitem[{{Miralles} {et~al.}(1997){Miralles}, {Salas}, {Cruz-Gonz{\'a}lez}, \&
  {Kurtz}}]{miralles1997}
{Miralles}, M.~P., {Salas}, L., {Cruz-Gonz{\'a}lez}, I., \& {Kurtz}, S. 1997,
  \apj, 488, 749

\bibitem[{{Moffat}(1969)}]{moffat1969}
{Moffat}, A.~F.~J. 1969, \aap, 3, 455

\bibitem[{{Ojha} {et~al.}(2011){Ojha}, {Samal}, {Pandey}, {Bhatt}, {Ghosh},
  {Sharma}, {Tamura}, {Mohan}, \& {Zinchenko}}]{ojha2011}
{Ojha}, D.~K., {Samal}, M.~R., {Pandey}, A.~K., {et~al.} 2011, \apj, 738, 156

\bibitem[{Pinna {et~al.}(2016)Pinna, Esposito, Hinz, Agapito, Bonaglia,
  Puglisi, Xompero, Riccardi, Briguglio, Arcidiacono, Carbonaro, Fini, Montoya,
  \& Durney}]{pinna2016}
Pinna, E., Esposito, S., Hinz, P., {et~al.} 2016, in Adaptive Optics Systems V,
  ed. E.~Marchetti, L.~M. Close, \& J.-P. V{\'e}ran, Vol. 9909, International
  Society for Optics and Photonics (SPIE), 99093V

\bibitem[{{Snell} \& {Bally}(1986)}]{snell1986}
{Snell}, R.~L. \& {Bally}, J. 1986, \apj, 303, 683

\bibitem[{{Stamatellos} {et~al.}(2011){Stamatellos}, {Whitworth}, \&
  {Hubber}}]{stamatellos2011}
{Stamatellos}, D., {Whitworth}, A.~P., \& {Hubber}, D.~A. 2011, \apj, 730, 32

\bibitem[{{Stamatellos} {et~al.}(2012){Stamatellos}, {Whitworth}, \&
  {Hubber}}]{stamatellos2012}
{Stamatellos}, D., {Whitworth}, A.~P., \& {Hubber}, D.~A. 2012, \mnras, 427,
  1182

\bibitem[{{Stecklum} {et~al.}(2016){Stecklum}, {Caratti o Garatti}, {Cardenas},
  {Greiner}, {Kruehler}, {Klose}, \& {Eisloeffel}}]{stecklum2016}
{Stecklum}, B., {Caratti o Garatti}, A., {Cardenas}, M.~C., {et~al.} 2016, The
  Astronomer's Telegram, 8732, 1

\bibitem[{{Stecklum} {et~al.}(2021){Stecklum}, {Wolf}, {Linz}, {Caratti o
  Garatti}, {Schmidl}, {Klose}, {Eisl{\"o}ffel}, {Fischer}, {Brogan}, {Burns},
  {Bayandina}, {Cyganowski}, {Gurwell}, {Hunter}, {Hirano}, {Kim}, {MacLeod},
  {Menten}, {Olech}, {Orosz}, {Sobolev}, {Sridharan}, {Surcis}, {Sugiyama},
  {van der Walt}, {Volvach}, \& {Yonekura}}]{stecklum2021}
{Stecklum}, B., {Wolf}, V., {Linz}, H., {et~al.} 2021, \aap, 646, A161

\bibitem[{{Strom} {et~al.}(1976){Strom}, {Strom}, \& {Vrba}}]{strom1976}
{Strom}, K.~M., {Strom}, S.~E., \& {Vrba}, F.~J. 1976, \aj, 81, 308

\bibitem[{{Szymczak} \& {Kus}(2000)}]{szymczak2000}
{Szymczak}, M. \& {Kus}, A.~J. 2000, \aaps, 147, 181

\bibitem[{{Szymczak} {et~al.}(2018{\natexlab{a}}){Szymczak}, {Olech},
  {Sarniak}, {Wolak}, \& {Bartkiewicz}}]{szymczak2018}
{Szymczak}, M., {Olech}, M., {Sarniak}, R., {Wolak}, P., \& {Bartkiewicz}, A.
  2018{\natexlab{a}}, \mnras, 474, 219

\bibitem[{{Szymczak} {et~al.}(2018{\natexlab{b}}){Szymczak}, {Olech}, {Wolak},
  {G{\'e}rard}, \& {Bartkiewicz}}]{szymczak2018S255}
{Szymczak}, M., {Olech}, M., {Wolak}, P., {G{\'e}rard}, E., \& {Bartkiewicz},
  A. 2018{\natexlab{b}}, \aap, 617, A80

\bibitem[{{Szymczak} {et~al.}(2012){Szymczak}, {Wolak}, {Bartkiewicz}, \&
  {Borkowski}}]{szymczak2012}
{Szymczak}, M., {Wolak}, P., {Bartkiewicz}, A., \& {Borkowski}, K.~M. 2012,
  Astronomische Nachrichten, 333, 634

\bibitem[{{Tamura} {et~al.}(1991){Tamura}, {Gatley}, {Joyce}, {Ueno}, {Suto},
  \& {Sekiguchi}}]{tamura1991}
{Tamura}, M., {Gatley}, I., {Joyce}, R.~R., {et~al.} 1991, \apj, 378, 611

\bibitem[{{Tan} {et~al.}(2014){Tan}, {Beltr{\'a}n}, {Caselli}, {Fontani},
  {Fuente}, {Krumholz}, {McKee}, \& {Stolte}}]{tan2014}
{Tan}, J.~C., {Beltr{\'a}n}, M.~T., {Caselli}, P., {et~al.} 2014, Protostars
  and Planets VI, 149

\bibitem[{{Turner}(1971)}]{turner1971}
{Turner}, B.~E. 1971, \aplett, 8, 73

\bibitem[{{Uchiyama} {et~al.}(2020){Uchiyama}, {Yamashita}, {Sugiyama},
  {Nakaoka}, {Kawabata}, {Itoh}, {Yamanaka}, {Akitaya}, {Kawabata}, {Yonekura},
  {Saito}, {Motogi}, \& {Fujisawa}}]{uchiyama2020}
{Uchiyama}, M., {Yamashita}, T., {Sugiyama}, K., {et~al.} 2020, \pasj, 72, 4

\bibitem[{{Wang} {et~al.}(2010){Wang}, {Li}, {Abel}, \& {Nakamura}}]{wang2010}
{Wang}, P., {Li}, Z.-Y., {Abel}, T., \& {Nakamura}, F. 2010, \apj, 709, 27

\bibitem[{{Wang} {et~al.}(2011){Wang}, {Beuther}, {Bik}, {Vasyunina}, {Jiang},
  {Puga}, {Linz}, {Rod{\'o}n}, {Henning}, \& {Tamura}}]{wang2011}
{Wang}, Y., {Beuther}, H., {Bik}, A., {et~al.} 2011, \aap, 527, A32

\bibitem[{{Zakri} {et~al.}(2022){Zakri}, {Megeath}, {Fischer}, {Gutermuth},
  {Furlan}, {Hartmann}, {Karnath}, {Osorio}, {Safron}, {Stanke}, {Stutz},
  {Tobin}, {Allen}, {Federman}, {Habel}, {Manoj}, {Narang}, {Pokhrel},
  {Rebull}, {Sheehan}, \& {Watson}}]{zakri2022}
{Zakri}, W., {Megeath}, S.~T., {Fischer}, W.~J., {et~al.} 2022, \apjl, 924, L23

\bibitem[{{Zhu} {et~al.}(2009){Zhu}, {Hartmann}, \& {Gammie}}]{zhu2009}
{Zhu}, Z., {Hartmann}, L., \& {Gammie}, C. 2009, \apj, 694, 1045

\bibitem[{{Zinchenko} {et~al.}(2015){Zinchenko}, {Liu}, {Su}, {Salii},
  {Sobolev}, {Zemlyanukha}, {Beuther}, {Ojha}, {Samal}, \&
  {Wang}}]{zinchenko2015}
{Zinchenko}, I., {Liu}, S.~Y., {Su}, Y.~N., {et~al.} 2015, \apj, 810, 10

\bibitem[{{Zinchenko} {et~al.}(2020){Zinchenko}, {Liu}, {Su}, {Wang}, \&
  {Wang}}]{zinchenko2020}
{Zinchenko}, I.~I., {Liu}, S.-Y., {Su}, Y.-N., {Wang}, K.-S., \& {Wang}, Y.
  2020, \apj, 889, 43

\bibitem[{{Zinnecker} {et~al.}(1998){Zinnecker}, {McCaughrean}, \&
  {Rayner}}]{zinnecker1998}
{Zinnecker}, H., {McCaughrean}, M.~J., \& {Rayner}, J.~T. 1998, \nat, 394, 862

\end{thebibliography}

\begin{appendix}

\section{SOUL performance}\label{sect:appendix_soul}

\begin{figure*}[!htb]
        \includegraphics[width=0.95\textwidth]{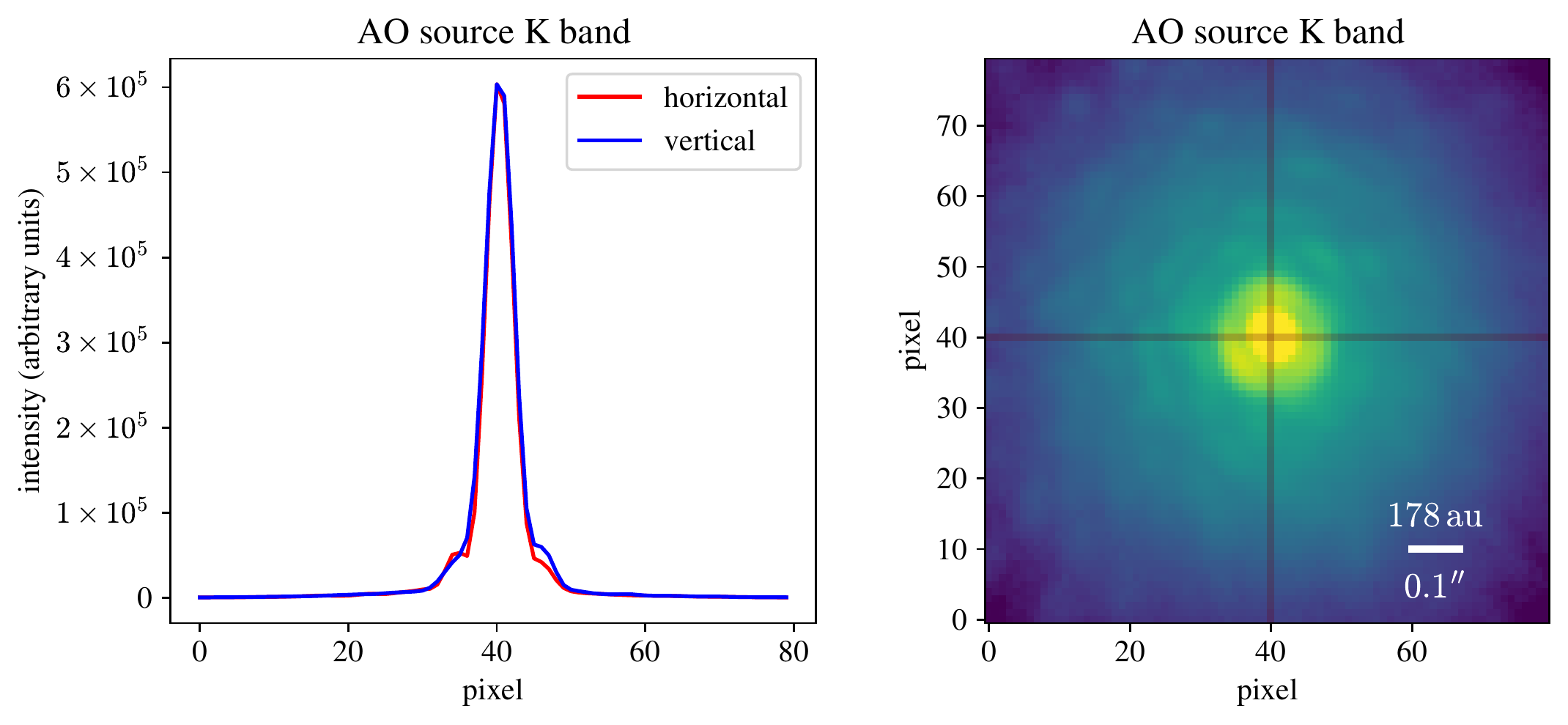}
        \includegraphics[width=0.95\textwidth]{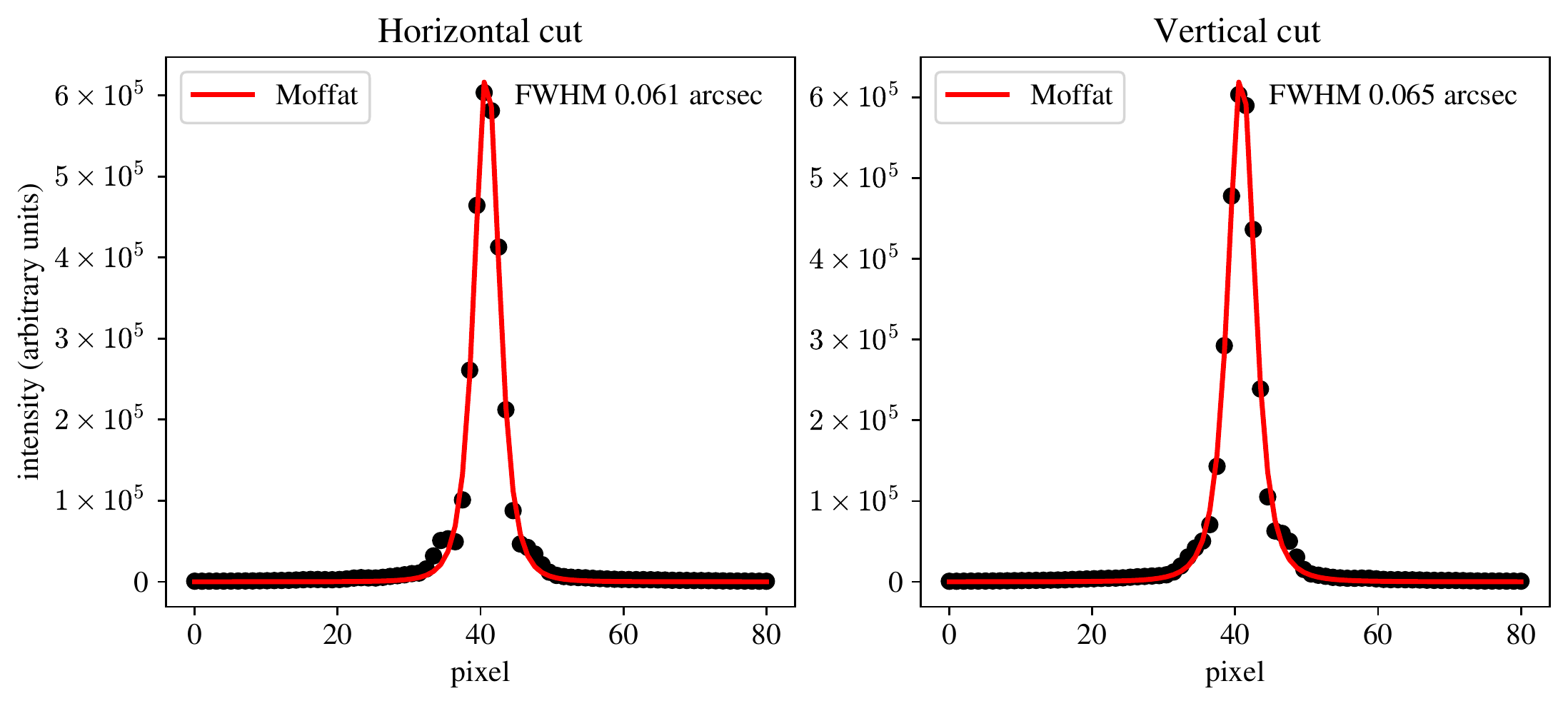}
  \caption{SOUL performance example featuring the K$_{\rm s}$ image with the AO guide star used in these observations. Top row: 1D profiles for the horizontal (red) and vertical (blue) cuts (left panel) as shown in the cutout image (right). Bottom row: Moffat fits to the horizontal (left panel) and the vertical (right panel) cuts, where the data is shown as black circles and the the best fit as red solid lines.}
     \label{fig:SOUL_performance}
\end{figure*}

To retrieve the FWHM in our images we fit a Moffat profile \citep{moffat1969} along the x and y axes of the image. The point spread function (PSF) under the assumption of circularity, i.e., azimuthally symmetric, is given by:

\begin{equation}
    f_{\rm Moffat}(r) = A\left[1+\left(\frac{r-r_0}{R}\right)^2\right]^{-\beta}  
\end{equation}

where $r$ is the distance from the centre $r_0$, $R$ is the core width and $\beta$ is the power. We fit the Moffat profile along the x and y directions for various sources. These sources are the AO guide star itself (2MASSJ06125505+1759289), the main science target (NIRS3), and the furthest star from the AO guide star common to the three filters. In Figure\,\ref{fig:SOUL_performance} top panels, we show the 1D intensity profile (left panel) and the 2D cutout (right panel) for the AO guide star. We extract these profiles at the peak intensity pixel. In the bottom panels, we show the Moffat fits as solid red lines. One can see that the Moffat profile fits very well the wings which are especially important in determining the shape of the PSF. To retrieve the FWHM in the Moffat profile we used the expression:

\begin{equation}
    {\rm FWHM_{Moffat}} = 2R\sqrt{2^{1/\beta}-1}
\end{equation}

In Table\,\ref{tab:soul_performance} we summarise the properties of three representative sources in our region as mentioned above. The reported columns are FWHM, central coordinates, and distance from the AO guide star as measured from the NIR images. We note that the FWHM is fairly constant among filters for the same source. However, one can see how the FWHM degrades with distance from the AO star. It is also worth noting that even though the NIRS3 region is not the furthest away from the AO guide star, it presents a larger FWHM than the furthest away star. This occurs because at this position the emission is dominated by extended nebulosity which increases the FWHM.

\begin{table*}
\caption{SOUL performance for three representative sources for the three filters used.}             % title of Table
\label{tab:soul_performance} 
\centering
\begin{tabular}{c c c c c c c}
\hline\hline
Source & Filter & FWHM$_x$ & FWHM$_y$ & RA(J2000) & Dec(J2000) & Distance from AO guide\\    % table heading 
 &  & (\arcsec) & (\arcsec) & (hh:mm:ss.ss) & (dd:mm:ss.ss) & (\arcsec)\\
\hline                        % inserts single horizontal line
    & K$_{\rm s}$ & 0.061 & 0.065 & 06:12:55.05 & 17:59:28.75 & 0 \\
   AO guide star & H$_2$ & 0.065 & 0.063 & & & \\
    & Br$\gamma$ & 0.063 & 0.064 & & &  \\
    & K$_{\rm s}$ & 0.080 & 0.075 & 06:12:54.01 & 17:59:23.04 & 16.0 \\
   NIRS3$^\dagger$ & H$_2$ & 0.096 & 0.080 & & & \\
    & Br$\gamma$ & 0.089 & 0.076 & & & \\
    & K$_{\rm s}$ & 0.067 & 0.075 & 06:12:54.20 & 17:59:11.38 & 21.38 \\
   Furthest star & H$_2$ & 0.067 & 0.075 & & & \\
    & Br$\gamma$ & 0.071 & 0.076 & & & \\    
\hline
\multicolumn{7}{l}{$^\dagger$ Note that this region is highly complex with a strong nebulous emission.}
\end{tabular}
\end{table*}

\clearpage
\section{Aperture Photometry}\label{sect:sedcreator}

\begin{figure}[!htb]
        \includegraphics[width=0.49\textwidth]{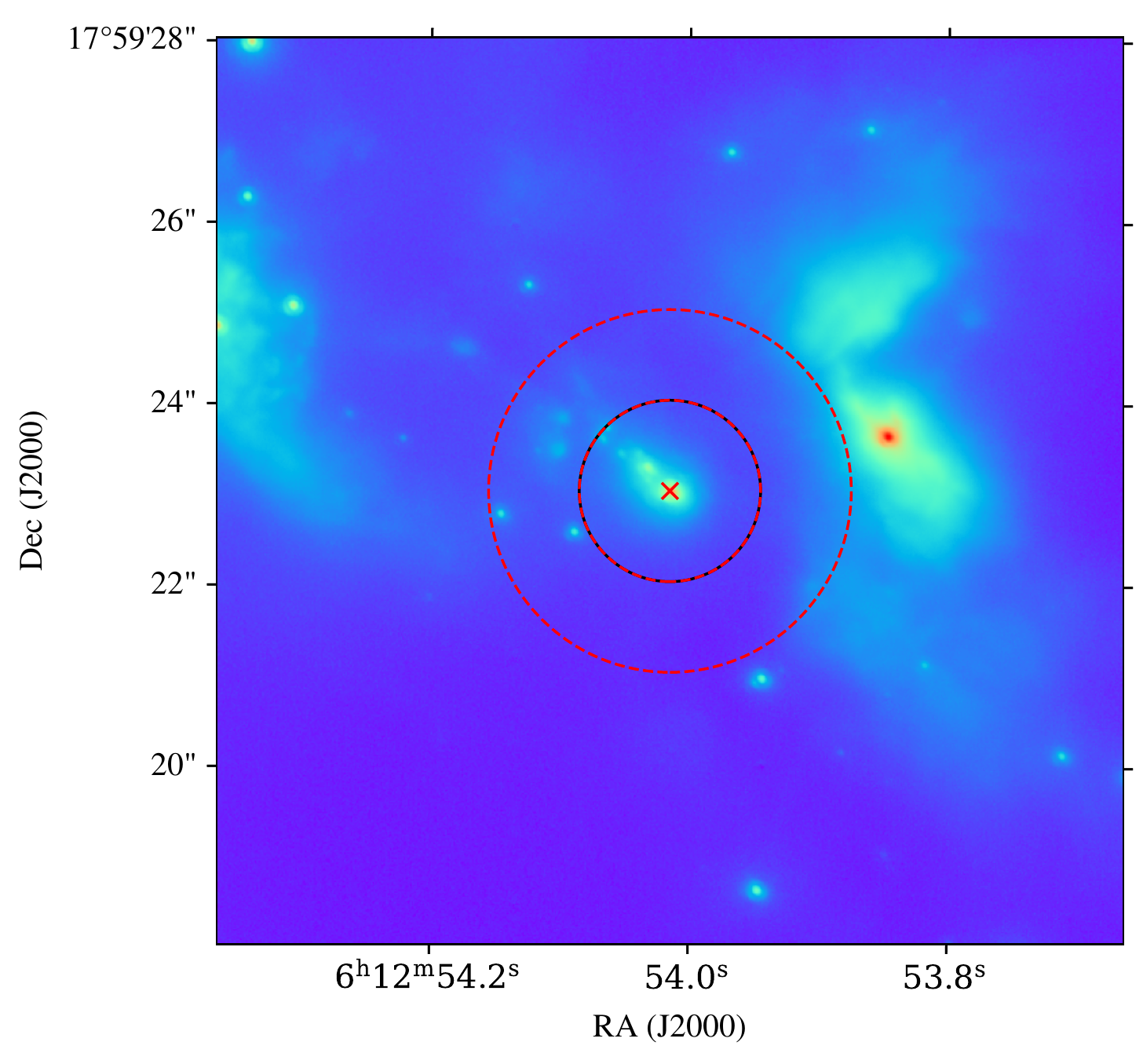}
        \includegraphics[width=0.49\textwidth]{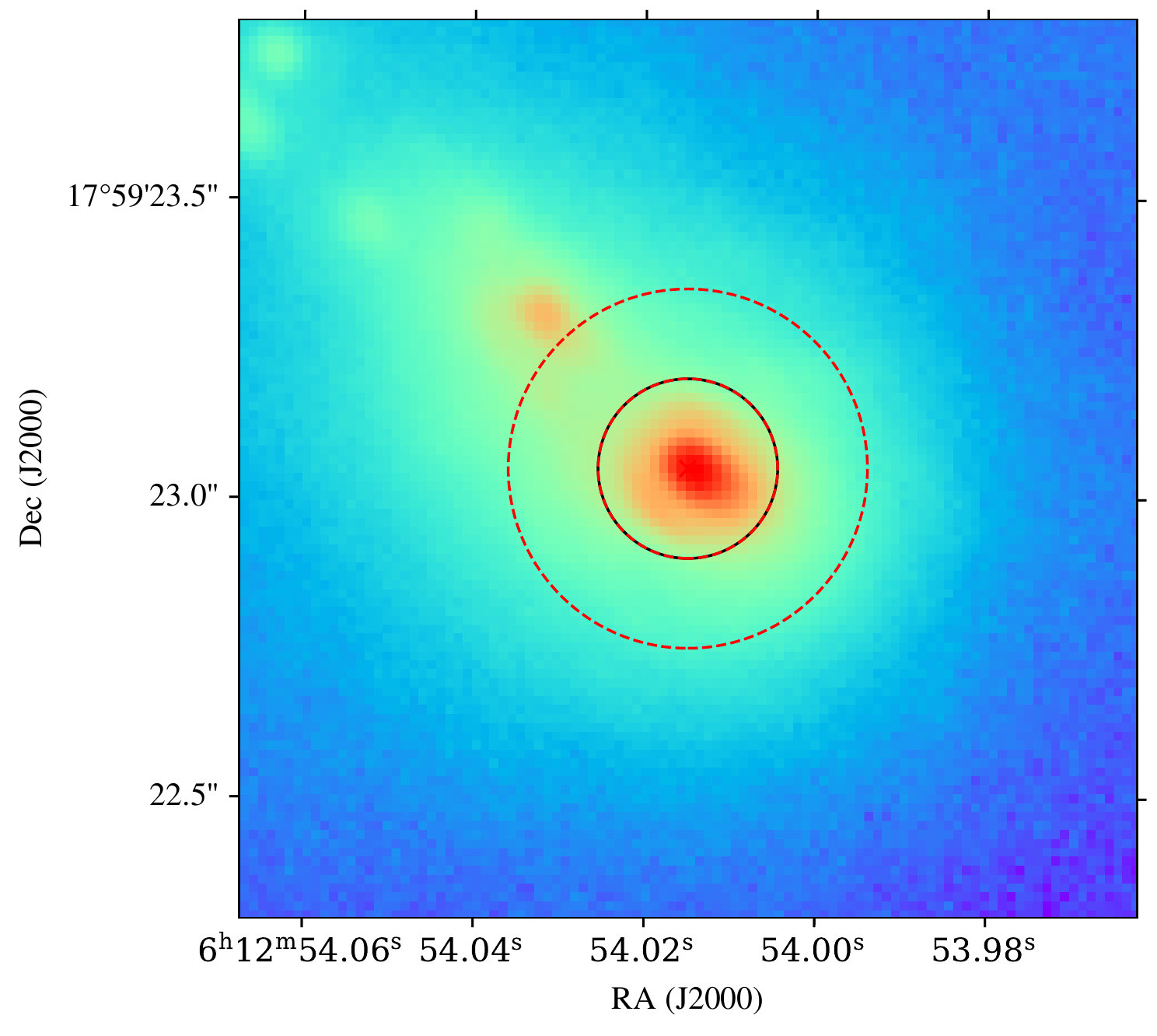}
  \caption{K$_{\rm s}$ band image zooming-in into the central region of S255IR. The red cross represents the centre of the aperture, whereas the black circle is the main aperture used, and the red dashed lines are the inner and outer radius of the annulus to account for background estimation. The aperture radius used was 1\arcsec (top panel) and 0\farcs15 (bottom panel).}
     \label{fig:aperture_phot_S255}
\end{figure}

We perform circular aperture photometry centering the main aperture in the central coordinates of NIRS3 (see Table\,\ref{tab:soul_performance}). We selected two apertures to perform two separate photometric measurements. The first with radius 1\arcsec, in order to compare with previous studies, and the second with radius 0\farcs15 in order to account for our higher angular resolution (best resolution in the NIR to date for NIRS3). We use the open source python package sedcreator \citep{fedriani2023} in particular the SedFluxer class to perform aperture photometry. Figure\,\ref{fig:aperture_phot_S255} shows the plot generated by the SedFluxer. The background is estimated from an annulus with inner and outer radii equal to 1 and 2 times the radius of the main aperture, represented as red dashed lines in the plots. The background is then considered to be the median value within the annulus multiplied by the area of the main aperture. This background value is then subtracted from the flux obtained in the main aperture. Table\,\ref{tab:photometry} summarises the 30 years of photometry.

\begin{table}[!h]
\caption{K band photometry for NIRS3 used in Figure\,\ref{fig:photometry_S255} and described in Section\,\ref{sect:photometry}.}             % title of Table
\label{tab:photometry} 
\centering
\begin{tabular}{r r c c}
\hline\hline
mjd & K & $^*$Resolution/ & Ref.\\    % table heading 
 &  & Aper. radius & \\    % table heading 
 (days) & (mag) & (\arcsec) & \\
\hline                        % inserts single horizontal line
    $47647.0$ & $10.10\pm0.10$ & 2.37 & \citet{tamura1991} \\
    $48237.0$ & $11.49\pm0.03$ & 2.0 & \citet{hodapp1994} \\
    $49411.0$ & $11.08\pm0.10$ & 1.5 & \citet{howard1997} \\
    $50115.0$ & $11.31\pm0.13$ & 1.15 & \citet{miralles1997} \\
    $51216.0$ & $12.56\pm0.00$ & 0.5-0.8 & \citet{bik2004thesis} \\
    $51584.0$ & $12.01\pm0.01$ & 1.32 & \citet{itoh2001} \\
    $51830.0$ & $11.82\pm0.02$ & 0.9 & \citet{ojha2011} \\
    $55173.0$ & $10.90\pm0.01$ & 1.0 & \citet{stecklum2016} \\
    $57354.0$ & $8.9\pm0.05$ & 2.5 & \citet{stecklum2016} \\
    $^\dagger57355.8$ & $8.68\pm0.02$ & 2.5 & \citet{uchiyama2020} \\
    $^\dagger57368.6$ & $8.03\pm0.05$ & 2.5 & \citet{uchiyama2020} \\
    $^\dagger58569.4$ & $11.75\pm0.05$ & 2.5 & \citet{uchiyama2020} \\
    $59623.0$ & $12.37\pm0.06$ & 1.0 & This work \\
\hline
\multicolumn{4}{l}{$^*$ We report the aperture radius used when available, otherwise}\\
\multicolumn{4}{l}{the resolution of the observations is reported.}\\
\multicolumn{4}{l}{$^\dagger$ We only report the first, maximum, and last photometric}\\
\multicolumn{4}{l}{points from \citet{uchiyama2020}, see their Table 1.}
\end{tabular}
\end{table}

\clearpage
\section{Non-annotated figures}

\begin{figure}[!htb]
        \includegraphics[width=0.49\textwidth]{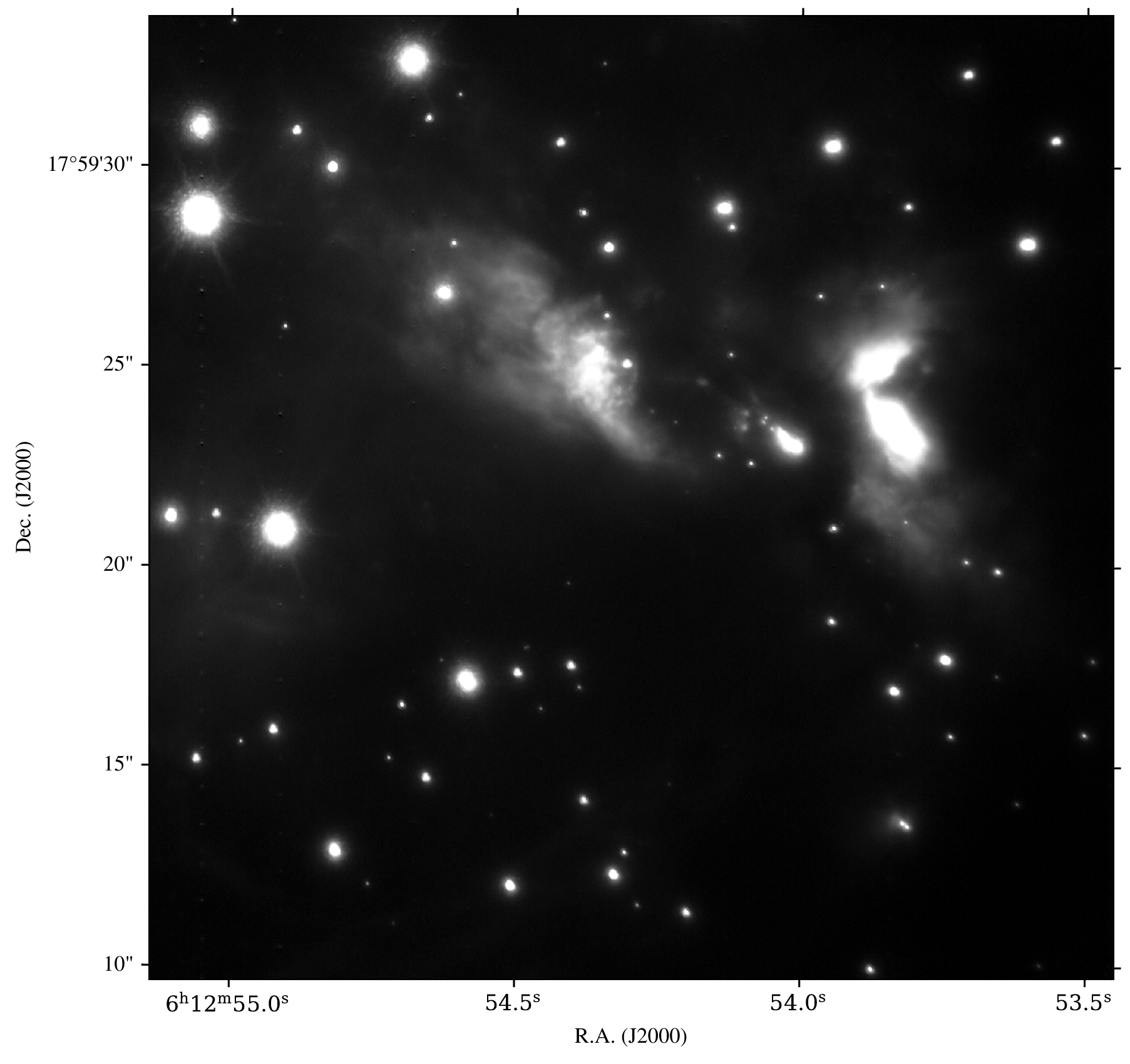}
        \includegraphics[width=0.49\textwidth]{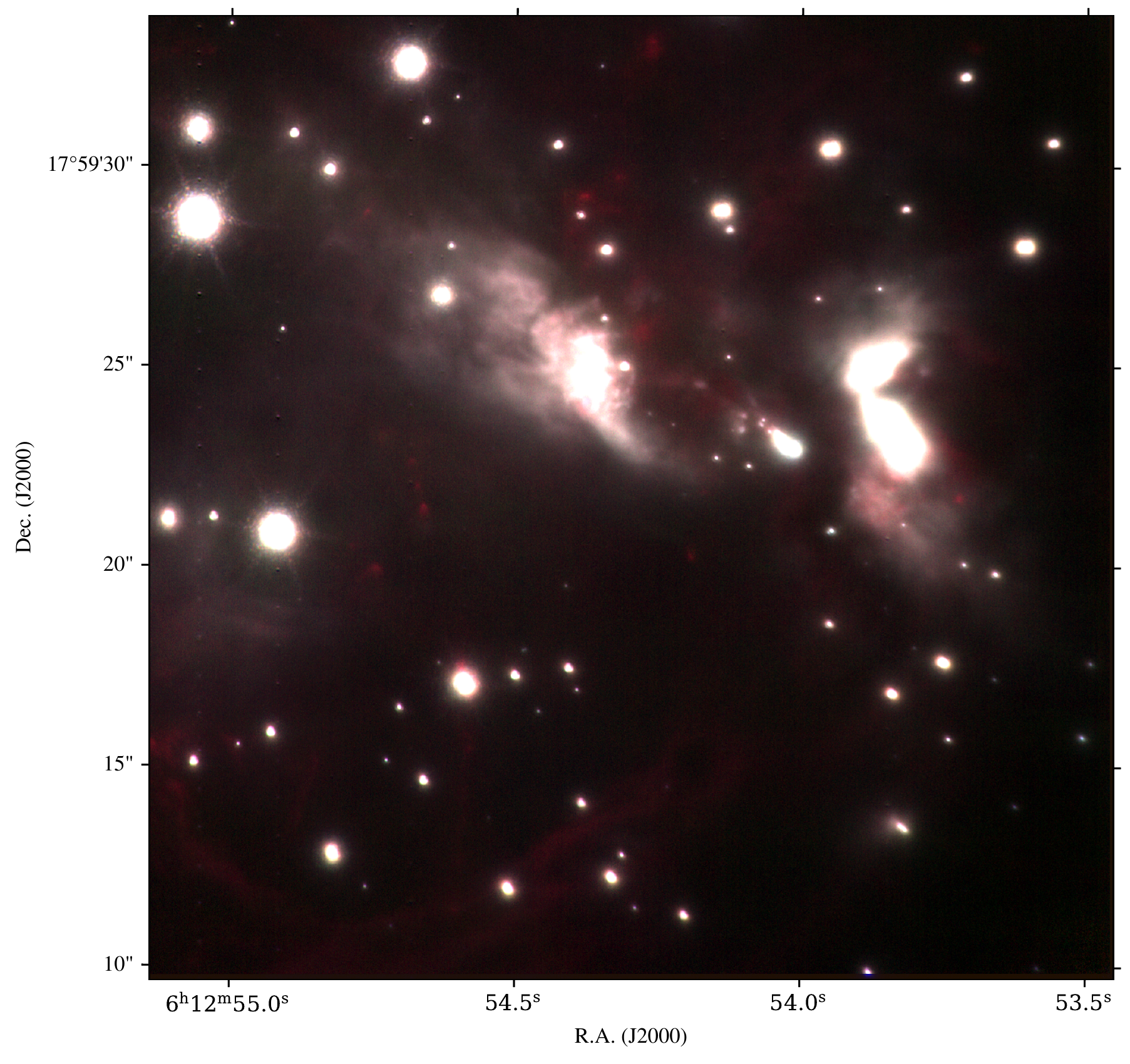}
  \caption{Same as Figure\,\ref{fig:continuum_image} but with no annotations.}
     \label{fig:continuum_image_non-annotated}
\end{figure}

\begin{figure}[!htb]
        \includegraphics[width=0.39\textwidth]{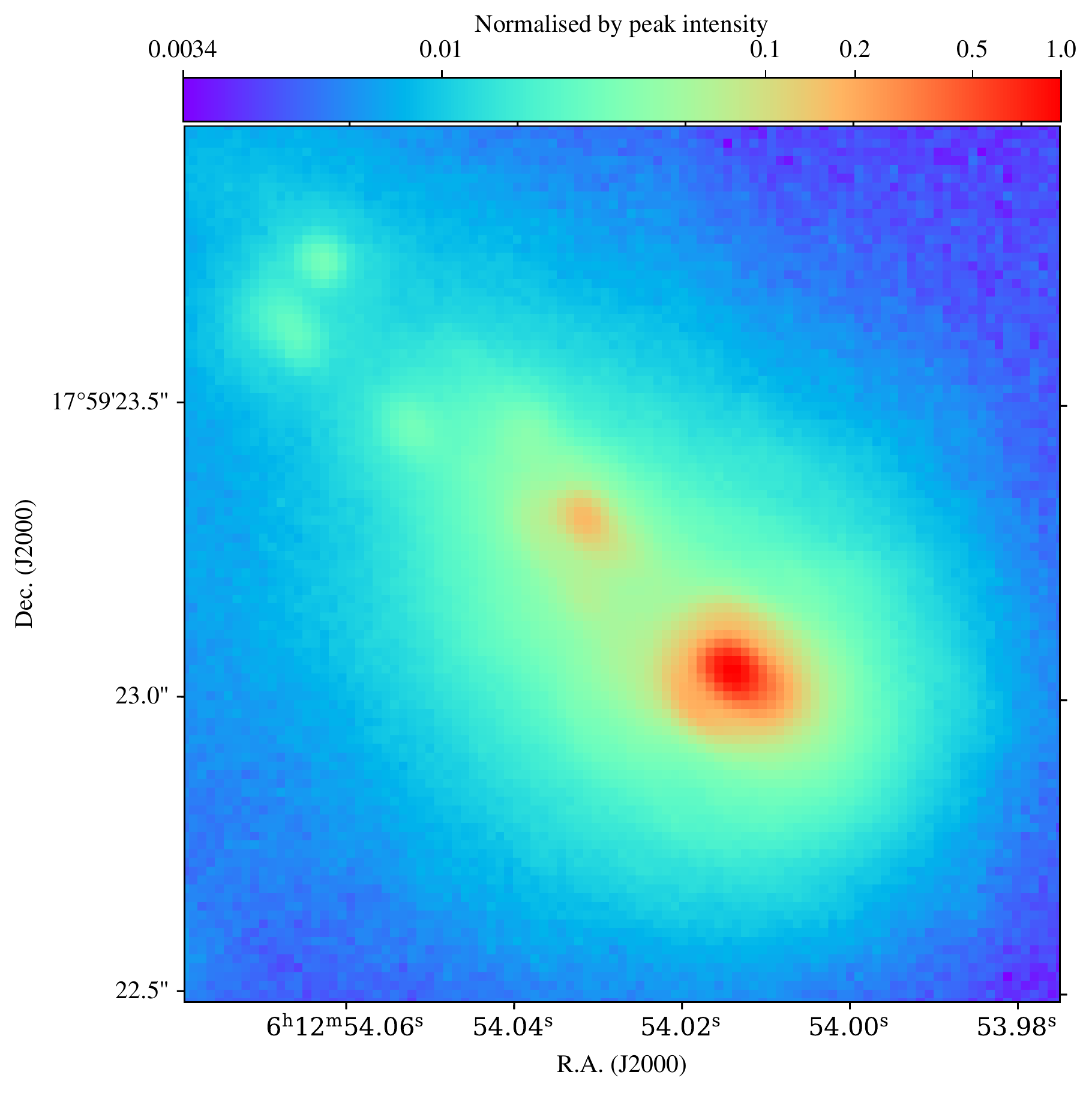}
  \caption{Same as Figure\,\ref{fig:S255_multiband} but with no annotations and no contours.}
     \label{fig:NEknot_non-annotated}
\end{figure}

\begin{figure}[!htb]
        \includegraphics[width=0.39\textwidth]{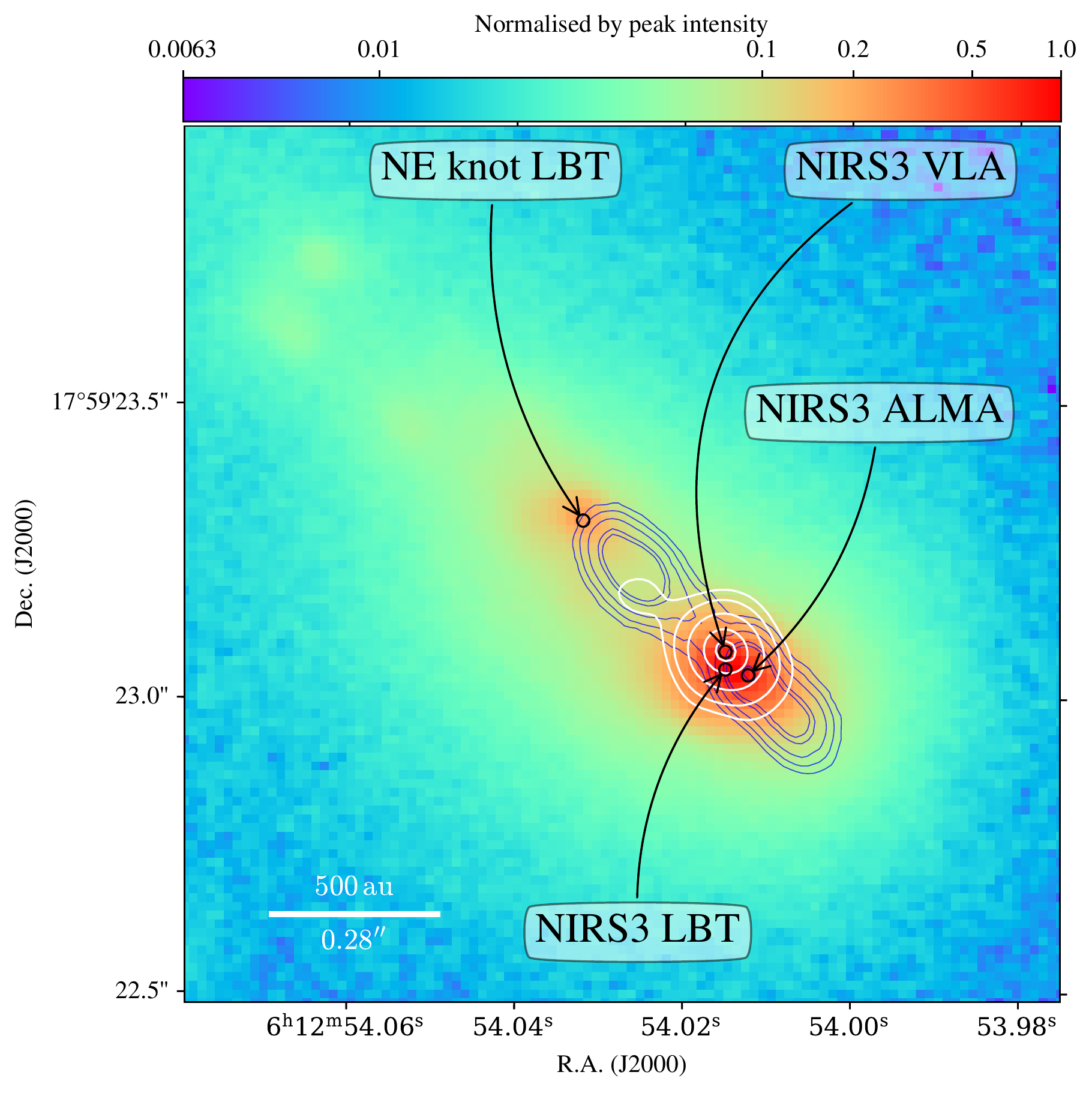}
        \includegraphics[width=0.39\textwidth]{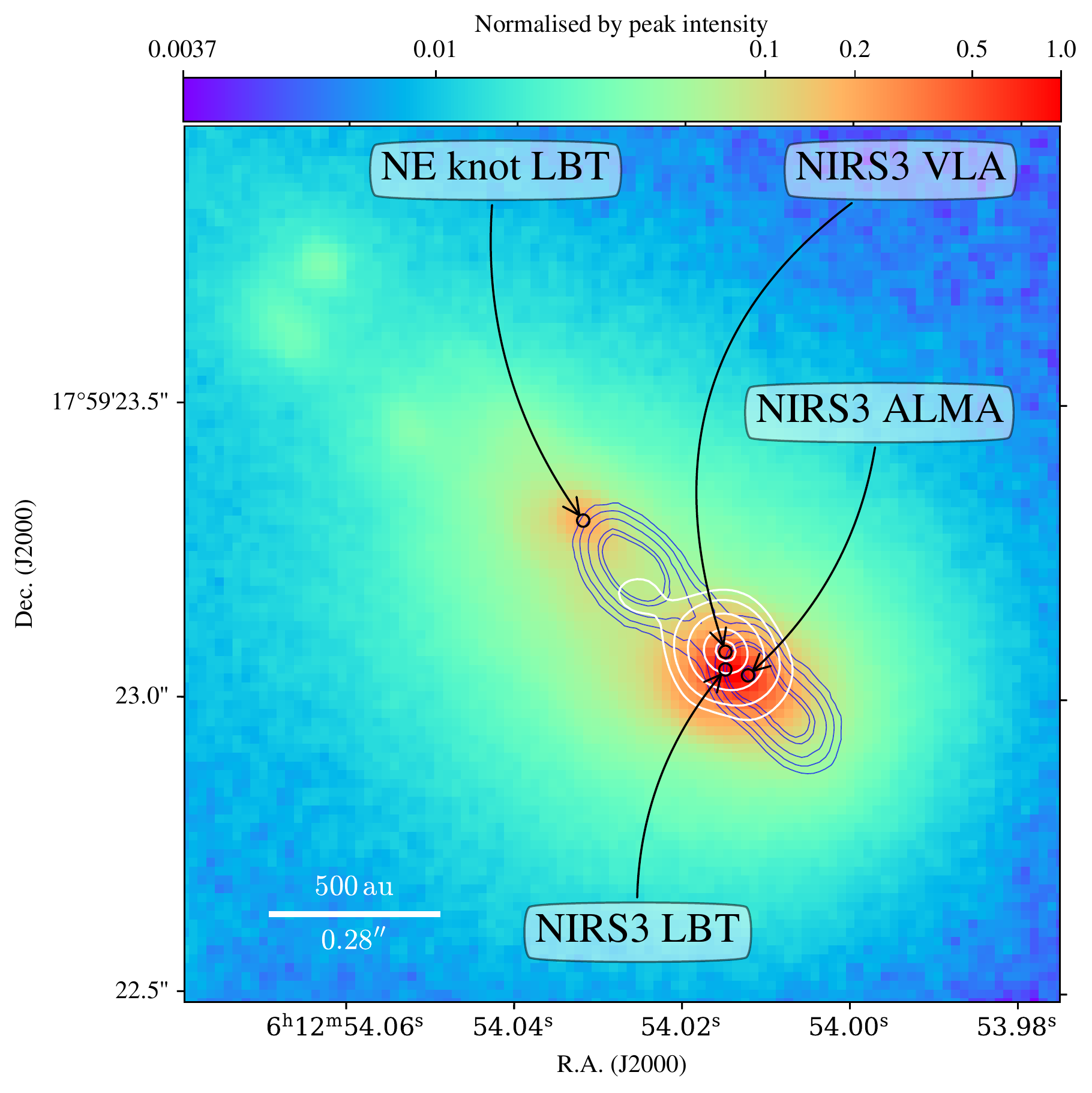}
  \caption{Same as Figure\,\ref{fig:S255_multiband} but for the filters H$_2$ (top panel) and Br$\gamma$ (bottom panel).}
     \label{fig:H2_and_BrG_images}
\end{figure}

In this section, we present the figures used in the main text removing the annotations, Figure\,\ref{fig:continuum_image_non-annotated}, and the contours, Figure\,\ref{fig:NEknot_non-annotated} to appreciate all the details. We also present the H$_2$ and Br$\gamma$ images in Figure\,\ref{fig:H2_and_BrG_images}.

\end{appendix}

\end{document}